%% file: template.tex
\documentclass{article}

\usepackage{arxiv}

\usepackage[utf8]{inputenc} % allow utf-8 input
\usepackage[T1]{fontenc}    % use 8-bit T1 fonts
\usepackage{hyperref}       % hyperlinks
\usepackage{url}            % simple URL typesetting
\usepackage{booktabs}       % professional-quality tables
\usepackage{amsfonts}       % blackboard math symbols
\usepackage{nicefrac}       % compact symbols for 1/2, etc.
\usepackage{microtype}      % microtypography
\usepackage{lipsum}         % Can be removed after putting your text content
\usepackage{amsmath}
\usepackage{cleveref}       % smart cross-referencing
\usepackage{graphicx}
\usepackage{subfig}
\usepackage{multicol}
\usepackage[sort&compress,square,numbers]{natbib}
\usepackage{doi}
\usepackage{xcolor}
\usepackage{appendix}
\DeclareMathOperator*{\argmax}{arg\,max}
\DeclareMathOperator*{\argmin}{arg\,min}

\title{The Automated Discovery of Kinetic Rate Models -- Methodological Frameworks}

\author{ \href{https://orcid.org/0000-0001-5273-7491}{\includegraphics[scale=0.06]{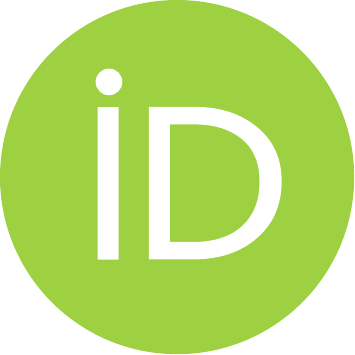}\hspace{1mm}Miguel Ángel de Carvalho Servia} \\
	Department of Chemical Engineering\\
	Imperial College London\\
    South Kensington, London, SW7 2AZ, UK\\
	\texttt{m.de-carvalho-servia21@imperial.ac.uk} \\
	\And
\href{https://orcid.org/0000-0001-9648-5265}{\includegraphics[scale=0.06]{orcid.pdf}\hspace{1mm}Ilya Orson Sandoval} \\
	Department of Chemical Engineering\\
	Imperial College London\\
    South Kensington, London, SW7 2AZ, UK\\
	\texttt{o.sandoval-cardenas20@imperial.ac.uk} \\
 \And
  \href{https://orcid.org/0000-0002-1163-0505}{\includegraphics[scale=0.06]{orcid.pdf}\hspace{1mm}King Kuok (Mimi) Hii} \\
	Department of Chemistry\\
	Imperial College London\\
    White City, London, W12 0BZ, UK\\
	\texttt{mimi.hii@imperial.ac.uk}
  \And
 \href{https://orcid.org/0000-0002-4630-1015}{\includegraphics[scale=0.06]{orcid.pdf}\hspace{1mm}Klaus Hellgardt} \\
	Department of Chemical Engineering\\
	Imperial College London\\
    South Kensington, London, SW7 2AZ, UK\\
	\texttt{k.hellgardt@imperial.ac.uk}
 \And
   \href{https://orcid.org/0000-0001-5956-4618}{\includegraphics[scale=0.06]{orcid.pdf}\hspace{1mm}Dongda Zhang $^*$} \\
    Department of Chemical Engineering\\
    The University of Manchester\\
    Manchester, M13 9PL, UK \\
    \texttt{dongda.zhang@manchester.ac.uk} 
  \And
  \href{https://orcid.org/0000-0003-0274-2852}{\includegraphics[scale=0.06]{orcid.pdf}\hspace{1mm}Ehecatl Antonio del Rio Chanona $^*$} \\
	Department of Chemical Engineering\\
	Imperial College London\\
    South Kensington, London, SW7 2AZ, UK\\
	\texttt{a.del-rio-chanona@imperial.ac.uk}
}

\renewcommand{\headeright}{}
\renewcommand{\undertitle}{}

\begin{document}
\maketitle

\begin{abstract}
\input{abstract.tex}
\end{abstract}

% keywords can be removed
\providecommand{\keyword}[1]{\textbf{Keywords:} #1}
\keyword{chemical reaction engineering, kinetic model generation, automated knowledge discovery, information criteria, symbolic regression}

\pagebreak

\section{Introduction}\label{Introduction}
\input{introduction.tex}

\section{Methodological Frameworks}\label{Methodological Frameworks}
\input{methodology.tex}

\section{Catalytic Kinetic Case Studies}\label{Case Studies}
\input{casestudies.tex}

\section{Results and Discussions}\label{Results and Discussions}
\input{results_n_discussion.tex}

\section{Conclusions}\label{Conclusions}
\input{conclusions.tex}

\section*{Declaration of Competing Interest}
The authors declare that they have no known competing financial interests or personal relationships that could have appeared to influence the work reported in this paper.

\section*{Acknowledgments and Funding}
This work was supported by the Engineering and Physical Sciences Research Council (EPSRC) funding grant EP/S023232/1.

\section*{Code and Data Availability}
The code used to produce all results and graphs shown in this work can be accessed at \url{https://github.com/MACServia/auto_discov_kin_rate_models}.

\clearpage
\begin{appendices}
\section{Case Studies}
\subsection{The Hypothetical Isomerization Reaction}\label{Isomerization}
The simplest case study used to benchmark the performance of the proposed methodologies, ADoK-S and ADoK-W, is a catalytic isomerization reaction, where \textit{A} is transformed to \textit{B} reversibly over a catalytic active site. The reaction is shown below:

\begin{align}\label{eq:5.1}
    A \rightleftharpoons B.
\end{align}

The kinetic rate model that describes the evolution of the concentrations of \textit{A} and \textit{B} through time is shown below. This expression has been directly borrowed from \citet{Marin_2019}.

\begin{align}\label{eq:5.2}
    r=-\frac{dC_A}{dt}=\frac{dC_B}{dt}=\frac{K_AC_A-K_BC_B}{K_CC_A+K_DC_B+K_E}
\end{align}

In Eq. \eqref{eq:5.2}, $C_A$ and $C_B$ represent the concentration of reactant \textit{A} and product \textit{B}, respectively. The kinetic parameters of the kinetic rate model are represented by $K_i$ where $i \in [A, B, ..., E]$. To generate the necessary data set to test both frameworks, five computational experiments are carried out, each with different initial conditions. The computational experiments are run with the following initial conditions (in molar units, mol L$^{-1}$, M): $(C_{A}(t=0), C_{B}(t=0)) \in \{(2, 0), (10, 2), (2, 2), (10, 2), (10, 1)\}$. For each computational experiment, the concentration of the reactant and product are recorded 30 times, at evenly spaced intervals between time $t_0=0$ h and $t_f=10$ h. For this particular case study, the kinetic parameters were defined as: $K_A=7$ M h$^{-2}$, $K_B=3$ M h$^{-2}$, $K_C=4$ h$^{-1}$, $K_D=2$ h$^{-1}$ and $K_E=6$ M h$^{-1}$.

Gaussian noise is added to the in-silico measurements to simulate a realistic chemical system. The added noise had zero mean and a standard deviation of 0.2 for both \textit{A} and \textit{B}. This noise addition allows the approximation of the response of a real system.

\subsection{The Decomposition of Nitrous Oxide}\label{Decomposition}
The second case study used for the performance analysis of ADoK-S and ADoK-W is the catalytic decomposition of nitrous oxide, where nitrous oxide ($N_2O$) is transformed to nitrogen gas ($N_2$) and oxygen gas ($O_2$). The reaction is shown below:

\begin{align}\label{eq:5.3}
    2N_2O \rightleftharpoons 2N_2 + O_2.
\end{align}

The kinetic rate model that describes the evolution of the concentrations of $N_2O$, $N_2$ and $O_2$ through time is shown below. This expression has been directly borrowed from \citet{Levenspiel_1998}.

\begin{align}\label{eq:5.4}
    r&=-2\frac{dC_{N_2O}}{dt}=2\frac{dC_{N_2}}{dt}=\frac{dC_{O_2}}{dt} =\frac{K_AC_{N_2O}^2}{1+K_BC_{N_2O}}
\end{align}

In Eq. \eqref{eq:5.4}, $C_{N_2O}$, $C_{N_2}$ and $C_{O_2}$ represent the concentration of reactant nitrous oxide, and of products nitrogen gas and oxygen gas, respectively. The kinetic parameters of the kinetic rate model are represented by $K_i$ where $i \in [A, B]$. To generate the necessary data set to test both frameworks, five computational experiments are carried out, each with different initial conditions. The experiments are run with the following initial conditions (in molar units): $(C_{N_2O}(t=0), C_{N_2}(t=0), C_{O_2}(t=0)) \in \{(5, 0, 0), (10, 0, 0), (5, 2, 0), (5, 0, 3), (0, 2, 3)\}$. For each experiment, the concentration of the reactant and products are recorded 30 times, at evenly spaced intervals between time $t_0=0$ h and $t_f=10$ h. For this particular case study, the kinetic parameters were defined as: $K_A=2$ M$^{-1}$ h$^{-1}$ and $K_B=5$ M$^{-1}$.

In-silico data sets are corrupted with Gaussian noise to emulate the behavior of actual chemical systems, resulting in the desired dynamic concentration trajectories. The introduced noise, characterized by a zero mean and a standard deviation of 0.2 for $N_2O$, $N_2$ and $O_2$, simulates the variability inherent in real system responses.

\subsection{The Hydrodealkylation of Toluene}\label{Hydrodealkylation_2}
The third case study used for the performance analysis of ADoK-S and ADoK-W is the catalytic toluene hydrodealkylation to benzene, where toluene ($C_6H_5CH_3$) and hydrogen gas ($H_2$) is transformed to benzene ($C_6H_6$) and methane ($CH_4$). The reaction is shown below:

\begin{align}\label{eq:5.5}
    C_6H_5CH_3 + H_2 \rightleftharpoons C_6H_6 + CH_4.
\end{align}

The kinetic rate model that describes the evolution of the concentrations of $C_6H_5CH_3$, $H_2$, $C_6H_6$ and $CH_4$ through time is shown below. This expression has been directly borrowed from \citet{Fogler_2016}.

\begin{align}\label{eq:5.6}
    r&=-\frac{dC_{T}}{dt}=-\frac{dC_{H}}{dt}=\frac{dC_{B}}{dt}=\frac{dC_{M}}{dt} =\frac{K_AC_TC_H}{1+K_BC_B+K_CC_T}
\end{align}

In Equation \ref{eq:5.6}, $C_{T}$, $C_{H}$, $C_{B}$ and $C_{M}$ represent the concentration of reactants toluene and hydrogen, and of products benzene and methane, respectively. The kinetic parameters of the kinetic rate model are represented by $K_i$ where $i \in [A, B, C]$. 

The computational experiments are run with the following initial conditions (in molar units): $(C_{T}(t=0), C_{H}(t=0), C_{B}(t=0), C_{M}(t=0)) \in \{(1, 8, 2, 3), (5, 8, 0, 0.5), (5, 3, 0, 0.5), (1, 3, 0, 3), (1, 8, 2, 0.5)\}$. For each experiment, the concentration of the reactant and products are recorded 30 times, at evenly spaced intervals between time $t_0=0$ h and $t_f=10$ h. The kinetic parameters were defined as: $K_A=2$ M$^{-1}$ h$^{-1}$, $K_B=9$ M$^{-1}$ and $K_C=5$ M$^{-1}$. 

Gaussian noise is added to the in-silico measurements to simulate a realistic chemical system. The added noise had zero mean and a standard deviation of 0.2 for $T$, $H$, $B$, $M$. This noise addition allows the approximation of the response of a real system. 

\section{Model Selection Analysis}\label{Model Selection Analysis}
Model selection is a pivotal aspect of the proposed methodologies, determining the most suitable model from an array of generated and optimized models. Consequently, an in-depth analysis on the behavior of four different information criteria is conducted: Akaike information criterion (AIC), sample corrected Akaike information criterion (AIC$_c$), Hannan-Quinn criterion (HQC) and Bayesian information criterion (BIC). Each criteria value for a given model $m$ is calculated using the equations presented below:

\begin{subequations}\label{eq:criteria}
\begin{gather}
    \text{AIC}_m = 2\mathcal{L}(\mathbf{\theta}_m\mid\mathcal{D})_m + 2d_m\\
    \text{AIC}_c,m = \text{AIC}_m + \frac{2(d_m+1)(d_m+2)}{n-d_m-2}\\
    \text{HQC}_m = 2\mathcal{L}(\mathbf{\theta}_m\mid\mathcal{D})_m + 2cd_m\log (\log (n))\\
    \text{BIC}_m = 2\mathcal{L}(\mathbf{\theta}_m\mid\mathcal{D})_m + d_m\textrm{log}(n),
\end{gather}
\end{subequations}

where $\mathcal{L}$ represents the negative log-likelihood (NLL), $\mathbf{\theta}_m$ are the parameters of model $m$, $\mathcal{D}$ represents the data set, $n$ represents the number of data points within set $\mathcal{D}$, and $d_m$ represents the number of parameters contained in $\mathbf{\theta}_m$. For HQC, $c$ stands for any constant equal to or greater than 1 to ensure model selection consistency (i.e., the best model within a model set is selected with probability going to one if the number of samples tend to infinity).

The analysis employs the hypothetical isomerization reaction detailed in Section \ref{Isomerization}, mimicking the data generation protocol as outlined therein.

This study examines seven competing kinetic models, with $r_5$ representing the data-generating kinetic model and the desired selection target for the information criteria.

\begin{subequations}\label{eq:5.7}
\begin{gather}
    r_1=K_1C_A\\
    r_2=K_1C_A-K_2C_B\\
    r_3=\frac{K_1C_A-K_2C_B}{K_3C_A}\\
    r_4=\frac{K_1C_A-K_2C_B}{K_3C_A+K_4C_B}\\
    r_5=\frac{K_1C_A-K_2C_B}{K_3C_A+K_4C_B+K_5}\\
    r_6=\frac{K_1C_A^2-K_2C_B-K_3C_A}{K_4C_A+K_5C_B+K_6}\\
    r_7=\frac{K_1C_A^2-K_2C_B^2-K_3C_A-K_4C_B}{K_5C_A+K_6C_B+K_7}
\end{gather}
\end{subequations}

The study presented herein aims at analyzing the behavior of the presented information criteria with respect to the noise and size of a data set.

\subsection{Noise Dependency}
The initial focus of this study is the exploration of the noise level employed in the generation of the kinetic data set, how it influences the behavior of the information criteria, and its eventual effect on model selection. For this, the same five experimental points specified in Section \ref{Isomerization} are used to create 13 distinct kinetic data sets with varying degrees of Gaussian noise, dictated by the user-defined variance $\sigma^2$. The variance range explored is $\sigma^2 \in [0.04, 0.25]$, at equally spaced intervals.

\begin{figure}[htb!]
    \centering
    \includegraphics[width=1\textwidth]{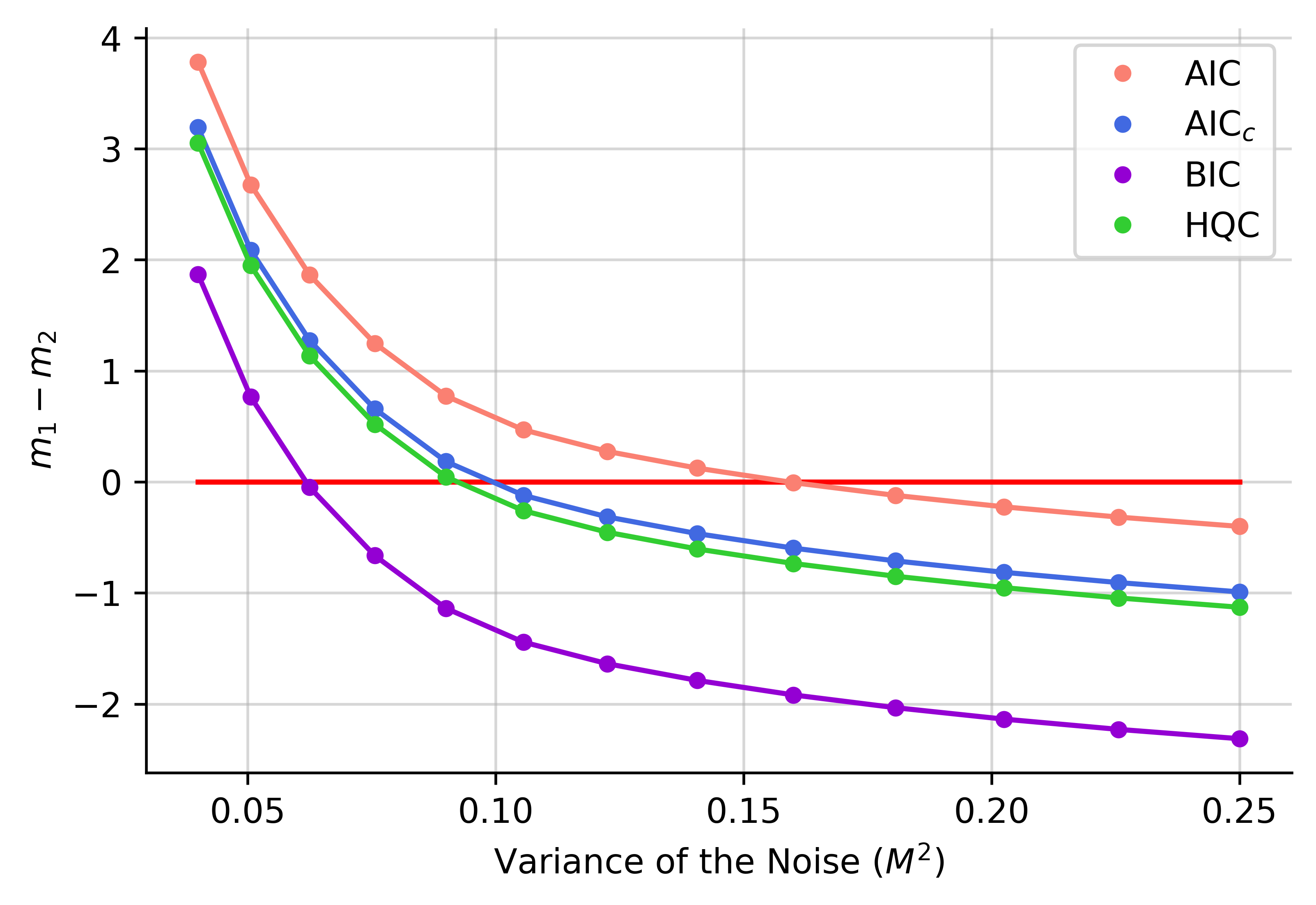}
    \caption{Plot of the difference of information criteria value between $m_1$ (the best model chosen from a subset that does not include the data-generating kinetic mode) and $m_2$ (the data-generating kinetic model) with respect to the variance of the Gaussian noise, $\sigma^2$, added to the kinetic data simulated, which was used to estimate the parameters of all rival models.}
    \label{fig:Noise Dependence}
\end{figure}

For each of these unique data sets, the data is utilized to re-calibrate the parameters for each of the seven candidate models, and subsequently the information criteria values are computed. Figure \ref{fig:Noise Dependence} presents a plot that illustrates the difference of information criteria value between the best kinetic model $m_1$ (chosen from a subset that excludes the data-generating model) and the data-generating model $m_2$. Within this graph, the horizontal line $y=0$ serves as the threshold at which an information criterion starts to select the incorrect model (i.e., above this line, the criterion selects the right model, below it, the criterion selects the wrong model).

It is worth noting that for each of the 13 sets of kinetic data, $m_1$ is consistently the 4-parameter model, $r_4$. Upon examining the graph, the AIC emerges as the most noise-resilient criterion, as its profile line intersects the horizontal threshold after all the other criteria (i.e., for a certain noise level, all other criteria select the wrong model, except AIC). Conversely, BIC exhibits the least noise resilience, as it initiates wrong model selection prior to the other criteria ($\sigma_i^2 \approx 0.06$). In general, a robustness hierarchy is preserved across these experiments, where $\textrm{AIC}>\textrm{AIC}\textrm{c}>\textrm{HQC}>\textrm{BIC}$ (from most to least robust).

The proposed hierarchy is a reasonable conclusion which can be deduced from mathematics. Given that the correct model is known to be $r_5$, that $m_1$ was invariably $r_4$, and that the number of samples remains constant ($n = 150$), the difference of the penalty imposed by each information criterion to both models, regardless of the data set, remains constant. More formally, where $d_m$ is the number of parameters of a model:

\begin{align}\label{New Eq}
    d_m \eta(d_m = 4, n = 150)_{\textrm{IC}} - d_m \eta(d_m = 5, n = 150)_{\textrm{IC}} = k_{IC},
\end{align}

where $\eta(d_m,n)$ is the penalty coefficient, with $\eta(d_m,n)_{\textrm{AIC}} = 2$, $\eta(d_m,n)_{\textrm{AIC}_\textrm{c}} = \frac{2n}{n-d_m-1}$, $\eta(d_m,n)_{\textrm{BIC}} = \log n$, and $\eta(d_m,n)_{\textrm{HQC}} = 2c\log \log n$. Eq. \eqref{New Eq} holds for each of the 13 kinetic data sets, where \textit{IC} stands for any of the four examined information criteria, and $k_{IC}$ represents an arbitrary constant. 

Considering the definitions of AIC, $\textrm{AIC}\textrm{c}$, BIC and HQC: $k_{\textrm{AIC}} = -2$, $k_{\textrm{AIC}\textrm{c}} = -2.14$, $k_{\textrm{BIC}} = -5.01$, $k_{\textrm{HQC}} = -3.22$. This demonstrates that AIC is the most tolerant criterion towards models of higher complexity (for AIC to favour $r_5$ over $r_4$, $2(l_{n = 150, m = 4} - l_{n = 150, m = 5}) > 2$; for BIC, $2(l_{n = 150, m = 4} - l_{n = 150, m = 5}) > 5.01$). Consequently, the aforementioned hierarchy is not only comprehensible but is also mathematically grounded.

However, the fact that all information criteria begin to choose the incorrect model at a certain noise level offers interesting insights. As previously explained, the penalty term's difference between the models $r_4$ and $r_5$ stays fixed across all data sets, hence the only element potentially influencing a selection shift is the NLL term. As the additive noise increases and the number of samples remains unchanged, the NLL values must also increase. Furthermore, deducing from Figure \ref{fig:Noise Dependence}, not only do NLL values increase with noise, but they also increase at different rates for each model. 

It becomes evident that the NLL term for the 5-parameter model rises at a higher rate in relation to the noise variance than the NLL term for the 4-parameter model (i.e., $\frac{\textrm{d}l_{n = 150, m = 5}}{\textrm{d}\sigma^2} > \frac{\textrm{d}l_{n = 120, m = 4}}{\textrm{d}\sigma^2}$) causing the criteria to select the wrong model more confidently as the noise is increased. It is important to underscore that this insight cannot be generalized to all 5-parameter and 4-parameter models, as it remains specific to this particular case.

\subsection{Quantity of Data Dependency}
\begin{figure}[htb!]
    \centering
    \includegraphics[width=\textwidth]{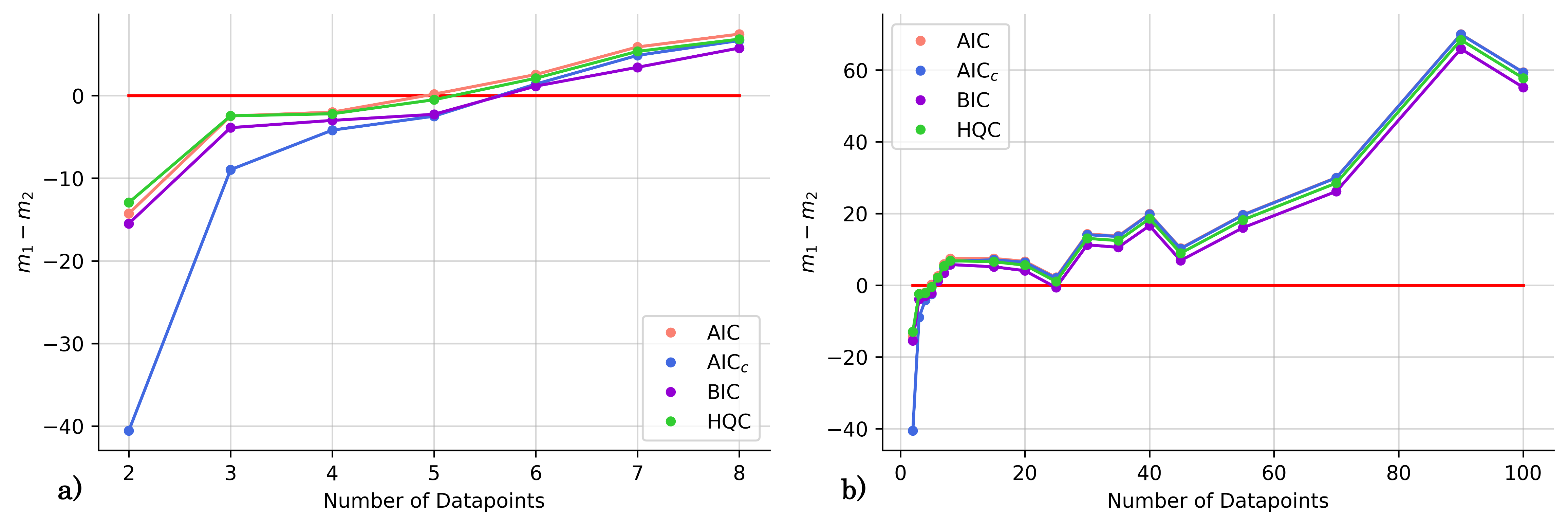}
    \caption{Plot of the difference between $m_1$ (the best model chosen from a subset that does not include the data-generating kinetic mode) and $m_2$ (the data-generating kinetic model) with respect to the number of data points available, which were used to estimate the parameters of all rival models. \textbf{a)}: shows a more detailed version of the graph in the low-data regime. \textbf{b)}: provides a more holistic perspective regarding the trends of the criteria in the high-data regime.}
    \label{fig:data point Dependence}
\end{figure}

The next aspect scrutinized is the influence of the number of samples on the values calculated by the information criteria. To investigate this, 18 data sets with varying numbers of data points are generated (i.e., the same five experiments detailed in Section \ref{Isomerization} are simulated, but with different number of samples per experiment). Gaussian noise, with a variance of $\sigma^2 = 0.2$, is introduced into the kinetic simulations. The outcomes of these computational experiments are presented in Figure \ref{fig:data point Dependence}.

A noteworthy feature from the graph deserving of discussion is how some criterion profiles intersect one another at different points of the plot, a phenomenon not identified in the previously presented graph. In the low-data regime, the HQC criterion is closest to identifying the correct model, followed closely by AIC, BIC, and $\textrm{AIC}_c$, in that order. The previously proposed hierarchy does not hold here, as the penalty terms now significantly differ, given their dependency on the number of samples.

However, with a mere six samples, the first intersection becomes visible in Figure \ref{fig:data point Dependence} \textbf{a)}, where $\textrm{AIC}_c$ starts selecting the data-generating kinetic model with greater confidence than the BIC, although both are still selecting the correct model. With approximately ten data points, a second intersection emerges in Figure \ref{fig:data point Dependence} \textbf{b)}, with $\textrm{AIC}_c$ now choosing the data-generating kinetic model with more certainty than the HQC, albeit the HQC is still selecting the right model. After this point, the original hierarchy reappears and is respected ($\textrm{AIC}>\textrm{AIC}_c>\textrm{HQC}>\textrm{BIC}$).

This finding is once again grounded by mathematics. As the number of data points increases, the penalty terms for HQC and BIC also increase, while the penalty term for $\textrm{AIC}_c$ decreases and that for AIC remains constant. The penalty term for $\textrm{AIC}_c$ asymptotically approaches that of the AIC, and BIC's penalty term increases at a higher rate than that of HQC as the number of data points increases.

It is evident that with the collection of more process information, all information criteria are capable of identifying the correct model (for this model set), underscoring the importance of sufficient data for robust model structure selection. It is vital to acknowledge the apparent ``noise" in Figure \ref{fig:data point Dependence}, which stems from the dynamic definition of $m_1$. While in the noise effect study $m_1$ is always $r_4$, this phenomenon does not hold in this study, leading to the changing identity of $m_1$.

\subsection{Summary}
In this investigation, utilizing a simple isomerization case study, the behavior of several information criteria was dissected. Comprehensive analyses of how various factors, including inherent noise in the data set and the size of the data set, influence the performance of these criteria were provided. Every conclusion was explained and rooted in the corresponding mathematical form of these criteria. All findings seem to suggest a specific ranking for the information criteria, with the best criterion listed first:

\begin{itemize}
\item Akaike Information Criterion,
\item Sample Corrected Akaike Information Criterion,
\item Hannan-Quinn Criterion,
\item Bayesian Information Criterion.
\end{itemize}

It is crucial, however, to acknowledge that these findings are context-specific and may not universally apply across different case studies. Furthermore, it is worth noting (and remembering) that the choice of a model selection criterion can transcend into a philosophical discussion as different information criteria bring diverse philosophical assumptions to their derivation, and all address slightly different questions. Thus, while the presented ranking can serve as a helpful guide -- and it is in fact used in the methodological frameworks proposed -- it should not be misconstrued as an absolute measure of the performance of the discussed information criteria across all disciplines.

\section{Model Discrimination}\label{Discrimination}
Model-based design of experiments (MBDoE) is critical in model discrimination, especially when the experimental budget has not yet been spent. For the methodological frameworks proposed herein, the Hunter-Reiner criterion has been adopted primarily because the evaluation of parameter uncertainty falls beyond the scope of this study, thus excluding any criteria necessitating the estimation of model response uncertainty at any given experimental point. Furthermore, the interpretability of Akaike weights design criterion is not of particular interest at this stage. The primary function of the Hunter-Reiner criterion is to identify the optimal experiment that maximizes the difference between the responses of two models.

In an ideal world, the primary goal of model discovery -- the central objective of this research -- is to pinpoint the optimal experiment that yields the largest discrepancy between a proposed model's and the data-generating model's response. Nevertheless, the actual underlying model is unknown, thus necessitating an approximation. This study examines two modeling approaches geared towards approximating the behavior of the data-generating kinetic model: Gaussian process state space model (GPSSM) and the second-best model generated from ADoK-S, hereinafter referred to as $\textrm{ADoK}_2$.

A GPSSM describes a nonlinear dynamical system, in which Gaussian processes are used to predict the state space dynamical transitions of a system (e.g., a reactive system) \cite{Frigola_2014, Eleftheriadis_2017}. This model comprises of a non-parametric representation of the system's dynamics rooted in Bayesian principles, complemented by hyperparameters that control the behavior of this non-parametric representation \cite{Frigola_2014}. GPSSMs have been chosen as one of the modeling approaches owing to their non-parametric nature, rendering them proficient in learning from small data sets (commonly found in kinetic studies), thereby outperforming parametric alternatives like recurrent neural networks \cite{Eleftheriadis_2017}. Even though GPSSMs are generally favored due to their probabilistic attributes (i.e., their ability to account for prediction uncertainty), this feature is not of particular relevance in this context, given that the Hunter-Reiner criterion does not factor in uncertainty. 

% While Gaussian processes are employed in this part of the research work and their regression is crucial, they do not constitute the primary focus of this work, thus pertinent information is referenced \cite{Rasmussen_2006, Wang_2019, Schulz_2018}. It is noteworthy that computing uncertainty for GPSSMs is fundamentally distinct from regular Gaussian processes, entailing a significantly more complex mathematical approach, as uncertainty must be propagated sequentially from one prediction to the next.

% To comprehend why uncertainty propagation is necessary, one should grasp the basic functioning of a GPSSM. Put simply, a GPSSM might intake a deterministic measurement of a state variable at time $t_0$ and output the forecasted state variables, accompanied by an approximation of uncertainty, at time $t_1$ (within the context of kinetic studies, these inputs and outputs correspond to the concentrations of observable species). For state predictions at $t_2$, the GPSSM employs its predictions at $t_1$ as an input. However, this input is probabilistic, not deterministic, consequently, it intuitively follows that the uncertainty at $t_i$ will always exceed the uncertainty at $t_{i+1}$, given that the calculation of the new prediction utilizes uncertain inputs. As such, uncertainty must be propagated.

% Although the GPSSMs' uncertainty is not computed in this study, for completeness, the reader is directed to the following manuscript \cite{Girard_2004}, where the process of uncertainty propagation in GPSSMs is explained in great detail. 

This self-contained study aims to comparatively evaluate the worst-case performance of the two models in approximating the data-generating model within a specified experimental space. The following procedure is employed for each of the three case studies used in this work (the hypothetical isomerization reaction, the decomposition of nitrous oxide, and the hydrodealkylation of toluene):

\begin{enumerate}
    \item Generate data corresponding to a given case study following the methodology detailed in the respective section;
    \item Normalize data to construct the training set for the GPSSM. The input training data set comprises of the concentrations of all observable species at $t \in [t_0, \cdots, t_{n-1}]$, while the output training data set consists of the concentrations of species \textit{X} at $t \in [t_1, \cdots, t_{n}]$; one GP is dedicated to each species observable in the reaction system;
    \item Train each GP using the compiled training data set and GPJax (a Python package which implements GPs using Jax) \cite{Pinder_2022};
    \item Execute one iteration of ADoK-S using the data sourced from the case studies and capture the second-best model generated, denoted as $\textrm{ADoK}_2$;
    \item Implement the Hunter-Reiner criterion to identify the experiment that maximizes the discrepancy between the data-generating model and the trained models (GPSSM and $\textrm{ADoK}_2$). This elucidates which model more accurately represents the real system in the worst-case scenario.
\end{enumerate}

Table \ref{Table: Model Discrimination} summarizes the sum of squared errors (SSE) between the response of the trained models and the data-generating model for the worst-case experiment. Based on these outcomes, it appears appropriate to choose the second-best kinetic model suggested by ADoK for implementing MBDoE alongside the best model. This approach enables further experiment generation when the modeler is not satisfied with the choices output by either ADoK-S or ADoK-W.

\begin{table}[htb!]
\caption{The worst-case scenario performance of the trained GPSSM and $\textrm{ADoK}_2$ with respect to the data-generating kinetic model for each case study.}
\begin{tabular}{lll}
\hline
         Case Study & SSE for GPSSM (M) & SSE for $\textrm{ADoK}_2$ (M) \\
\hline
    Hypothetical isomerization reaction &   0.27 &     0.11 \\
    Decomposition of nitrous oxide &    90.69 &     0.42 \\
    Hydrodealkylation of toluene &     49.20 &      0.14 \\
\hline
\end{tabular}
\label{Table: Model Discrimination}
    \centering
\end{table}

\section{Detailed ADoK-S Performance}\label{ADoK-S Performance_2}
\subsection{The Hypothetical Isomerization Reaction}\label{Iso Results}
Starting with five initial experiments, as delineated in Section \ref{Isomerization}, ADoK-S generated, optimized and selected the presented concentration profiles for each species and experiment. Below, $\hat{C}_{i,j}$ is the model that describes the dynamic evolution of the concentration of species $i$ during experiment $j$.

\begin{subequations}
\begin{gather}\label{eq:conc_iso_strong}
    \hat{C}_{A,1}(t) = \exp{(1.470 - \exp{(0.642t)})} + 0.634\\
    \hat{C}_{B,1}(t) = \frac{-1.325}{\exp{(t)}} + 1.376\\
    \hat{C}_{A,2}(t) = 0.075t^2 - 1.375t + 9.981\\
    \hat{C}_{B,2}(t) = t\exp{(-0.084t + 0.399)}\\
    \hat{C}_{A,3}(t) = \exp{(-0.615t)} + 1.213\\
    \hat{C}_{B,3}(t) = \exp{(1.040 - \exp{(-1.189 - t)})}\\
    \hat{C}_{A,4}(t) = \frac{9.918}{0.179t + 1.078} + 0.665\\
    \hat{C}_{B,4}(t) = -0.055t^2 + 1.128t + 2.087\\
    \hat{C}_{A,5}(t) = \frac{\exp{(\exp{(\exp{(-0.060t)})})}}{1.499}\\
    \hat{C}_{B,5}(t) = 0.608t + 2.069.
\end{gather}
\end{subequations}

Approximations of the rates were calculated by numerically differentiating the concentration profiles. These estimates, in turn, are used to generate rate models. The best two models, based on the AIC ranking, are presented below:

\begin{subequations}
\begin{gather}\label{eq:rate_iso_strong}
    \hat{r}_1(t) = \frac{C_A(C_A - 0.578)(C_A - 0.443)}{(C_A + 0.540)(0.684C_A(C_A - 0.578) + C_B(C_B - 0.949))}\\
    \hat{r}_2(t) = \frac{(C_A - 0.579)(C_A - 0.519)(C_A - 0.161)}{(C_A + 0.293)(C_B(C_B - 0.988) + 0.684(C_A - 0.519)(C_A - 0.161))}.
\end{gather}
\end{subequations}

With these two rate models, a new experiment was suggested using MBDoE. The proposed experiment is $(C_{A}(t=0), C_{B}(t=0)) = (4.926, 0.000)$ M. For the sixth experiment, ADoK-S generated, optimized and selected the presented concentration profiles.

\begin{subequations}
\begin{gather}\label{eq:conc_iso_strong_2}
    \hat{C}_{A,6}(t) = \exp{\Bigl(1.258 - \frac{t}{2.151}\Bigr)} + 1.458\\
    \hat{C}_{B,6}(t) = \frac{3.436}{\exp{(\exp{(-0.707t + 0.824)})}}.
\end{gather}
\end{subequations}

By approximating the rate measurements from the sixth experiment and concatenating the estimates to the previous data set, ADoK-S uncovers the structure and similar parameters to the data-generating rate model:

\begin{equation}
    \hat{r}^* = \frac{8.666C_A-3.642C_B}{4.976C_A+2.525C_B+7.003}.
\end{equation}

\subsection{The Decomposition of Nitrous Oxide}\label{Decom Results}
Starting with five initial experiments, as delineated in Section \ref{Decomposition}, ADoK-S generated, optimized and selected the presented concentration profiles for each species and experiment. Below, $\hat{C}_{i,j}$ is the model that describes the dynamic evolution of the concentration of species $i$ during experiment $j$

\begin{subequations}
\begin{gather}\label{eq:conc_decom_strong}
    \hat{C}_{N_2O,1}(t) = \exp{\Bigl(1.602 - \frac{t}{2.128}\Bigr)} + 0.309 \\
    \hat{C}_{N_2,1}(t) = 4.866 - \frac{4.703}{\exp{(0.383t)}} \\
    \hat{C}_{O_2,1}(t) = \frac{t}{0.347t + 0.896} + 0.152\\
    \hat{C}_{N_2O,2}(t) = \exp{\Bigl(2.287 - \frac{t}{2.486}\Bigr)} + 0.113 \\
    \hat{C}_{N_2,2}(t) = 9.863 - \frac{2.262}{\exp{(0.390t)}} \\
    \hat{C}_{O_2,2}(t) = t\exp{(\exp{(-0.094t)})} - t \\
    \hat{C}_{N_2O,3}(t) = \exp{(1.520 - 0.418t)} + 0.199 \\
    \hat{C}_{N_2,3}(t) = 6.970 - 5.032\exp{(-0.385t)} \\
    \hat{C}_{O_2,3}(t) = \frac{t}{0.318t + 0.850} \\
    \hat{C}_{N_2O,4}(t) = \exp{(1.533 - 0.380t)} + 0.187 \\
    \hat{C}_{N_2,4}(t) = 1.560t\exp{(-0.122t)} \\
    \hat{C}_{O_2,4}(t) = 5/343 - \exp{(1.741 - \exp{(0.270t)})} \\
    \hat{C}_{N_2O,5}(t) = 0.077 \\
    \hat{C}_{N_2,5}(t) = 2.009 \\
    \hat{C}_{O_2,5}(t) = 2.978.
\end{gather}
\end{subequations}

Approximations of the rates were calculated by numerically differentiating the concentration profiles. These estimates, in turn, are used to generate rate models. The best two models, based on the AIC ranking, are presented below:

\begin{subequations}
\begin{gather}\label{eq:rate_decom_strong}
    \hat{r}_1(t) = \frac{C_{N_2O}(0.416C_{N_2O} - 0.034)}{C_{N_2O} + 0.200}\\
    \hat{r}_2(t) = 0.061 - 0.404C_{N_2O}.
\end{gather}
\end{subequations}

With these two rate models, a new experiment was suggested using MBDoE. The proposed experiment is $(C_{N_2O}(t=0), C_{N_2}(t=0), C_{O_2}(t=0)) = (0.000, 1.641, 1.095)$ M. For the sixth experiment, ADoK-S generated, optimized and selected the presented concentration profiles.

\begin{subequations}
\begin{gather}\label{eq:conc_decom_strong_2}
    \hat{C}_{N_2O,6}(t) = 0.158 \\
    \hat{C}_{N_2,6}(t) = 1.631 \\
    \hat{C}_{O_2,6}(t) = 1.085.
\end{gather}
\end{subequations}

By approximating the rate measurements from the sixth experiment and concatenating the estimates to the previous data set, ADoK-S uncovers the structure and similar parameters to the data-generating rate model:

\begin{equation}
    \hat{r}^* = \frac{1.937C_{N_2O}^2}{1+4.803C_{N_2O}}.
\end{equation}

\subsection{The Hydrodealkylation of Toluene}
Starting with five initial experiments, as delineated in Section \ref{Hydrodealkylation_2}, ADoK-S generated, optimized and selected the presented concentration profiles for each species and experiment. Below, $\hat{C}_{i,j}$ is the model that describes the dynamic evolution of the concentration of species $i$ during experiment $j$

\begin{subequations}
\begin{gather}\label{eq:conc_hydro_strong}
    \hat{C}_{T,1}(t) = \exp{(0.164 - t)} + 0.102 \\
    \hat{C}_{H,1}(t) = 7.006 + \exp{(-0.534t)} \\
    \hat{C}_{B,1}(t) = 3.021 + \frac{-0.525}{0.430 + t} \\
    \hat{C}_{M,1}(t) = 3.940 - \exp{(-0.525(t + 0.115))} \\
    \hat{C}_{T,2}(t) = \frac{6.652}{1.172+\frac{t}{1.592}} - 0.407 \\
    \hat{C}_{H,2}(t) = \exp{(\exp{(-0.229(t - 2.379))})}+ 2.424 \\
    \hat{C}_{B,2}(t) = \frac{t}{0.437 + 0.175t} \\
    \hat{C}_{M,2}(t) = \exp{\Big(1.775 - \frac{1.599}{\exp{(-0.414) + t}}\Big)} \\
    \hat{C}_{T,3}(t) = 1.295 + \exp{(\exp{(-0.197(t - 1.305))})} \\
    \hat{C}_{H,3}(t) = \exp{(\exp{(-0.070t^2)})} - 0.323 \\
    \hat{C}_{B,3}(t) = \exp{\Big(\frac{\exp{(1.004)}}{-0.793 - t} + 1.192\Big)} \\
    \hat{C}_{M,3}(t) = \exp{\Big(\frac{-1.330}{\exp{(0.472t)}} + 1.115\Big)} \\
    \hat{C}_{T,4}(t) = \exp{(-0.151\exp{(t)})} + 0.058 \\
    \hat{C}_{H,4}(t) = \frac{-2.282}{-2.029 - t^2} + 1.948 \\
    \hat{C}_{B,4}(t) = \exp{\Big(\frac{-2.523}{\exp{(t)}}\Big)} \\
    \hat{C}_{M,4}(t) = 0.230t - 0.016t^2 + 3.318 \\
    \hat{C}_{T,5}(t) = \frac{t + 0.493}{\exp{(t)}} + 0.063 \\
    \hat{C}_{H,5}(t) = 6.906 + \exp{(-0.376t)} \\
    \hat{C}_{B,5}(t) = 3.010 - \exp{(-0.442t)} \\
    \hat{C}_{M,5}(t) = \frac{\frac{-0.797}{0.557 + t} + 1.574}{1.009}.
\end{gather}
\end{subequations}

Approximations of the rates were calculated by numerically differentiating the concentration profiles. These estimates, in turn, are used to generate rate models. The best two models, based on the AIC ranking, are presented below:

\begin{subequations}
\begin{gather}\label{eq:rate_hydro_strong}
    \hat{r}_1(t) = -0.052C_TC_H\\
    \hat{r}_2(t) = 0.020C_T - 0.056C_TC_H.
\end{gather}
\end{subequations}

With these two rate models, a new experiment was suggested using MBDoE. The proposed experiment is $(C_{T}(t=0), C_{H}(t=0), C_{B}(t=0), C_{M}(t=0)) = (1.948, 7.503, 1.232, 2.504)$ M. For the sixth experiment, ADoK-S generated, optimized and selected the presented concentration profiles.

\begin{subequations}
\begin{gather}\label{eq:conc_hydro_strong_2}
    \hat{C}_{T,6}(t) = \exp{\Bigl(\frac{1.216 - t}{\exp{(0.645)}}\Bigr)} \\
    \hat{C}_{H,6}(t) = \frac{4.934}{\exp{(t) + \exp{(0.330)}}} + 5.591 \\
    \hat{C}_{B,6}(t) = \frac{3.004}{\exp{(\exp{(-0.862t)})}} \\
    \hat{C}_{M,6}(t) = 4.896 + \frac{4.048}{-1.506 - t}.
\end{gather}
\end{subequations}

By approximating the rate measurements from the sixth experiment and concatenating the estimates to the previous data set, ADoK-S is still unable to uncover the correct model structure. The best two models from this iteration are presented below:

\begin{subequations}
\begin{gather}\label{eq:rate_hydro_strong_2}
    \hat{r}_1(t) = \frac{-1.000C_TC_H}{1.286C_T + 12.586}\\
    \hat{r}_2(t) = -0.034C_T(-0.741C_B^2 + 1.886C_B + 1.886C_H - C_T + 1.612) - 0.021.
\end{gather}
\end{subequations}

With these two rate models, a new experiment was suggested using MBDoE. The proposed experiment is $(C_{T}(t=0), C_{H}(t=0), C_{B}(t=0), C_{M}(t=0)) = (2.560, 5.654, 0.341, 2.337)$ M. For the seventh experiment, ADoK-S generated, optimized and selected the presented concentration profiles.

\begin{subequations}
\begin{gather}\label{eq:conc_hydro_strong_3}
    \hat{C}_{T,7}(t) = \exp{(\exp{(0.033 - t)})} - 0.101t \\
    \hat{C}_{H,7}(t) = \exp{(\exp{(-0.298t - 0.172)})} + 2.186 \\
    \hat{C}_{B,7}(t) = \exp{(1.024)} - \exp{(-0.442t + 0.802)} \\
    \hat{C}_{M,7}(t) = 4.818 - \exp{(-0.434(t - \exp{(0.693)}))}.
\end{gather}
\end{subequations}

By approximating the rate measurements from the seventh experiment and concatenating the estimates to the previous data set, ADoK-S uncovers the structure and similar parameters to the data-generating rate model:

\begin{equation}
    \hat{r}^* = \frac{1.669C_TC_H}{1+7.347C_B+4.439C_T}.
\end{equation}

\section{Detailed ADoK-W Performance}
\subsection{The Hypothetical Isomerization Reaction}\label{Iso Results W}
Starting from the initial five experiments, ADoK-W generated and optimized a plethora of rate models, outputting the best two models (based on AIC ranking) presented below:

\begin{subequations}
\begin{gather}\label{eq:rate_iso_weak}
    \hat{r}_1(t) = \frac{0.025C_A^2 + 0.967C_A - 0.597C_B + 0.438}{C_A}\\
    \hat{r}_2(t) = \frac{1.143 - 0.490_B}{C_A}.
\end{gather}
\end{subequations}

With these two rate models, a new experiment was suggested using MBDoE. The proposed experiment is $(C_{A}(t=0), C_{B}(t=0)) = (7.319, 2.000)$ M. After concatenating the measurements collected from the sixth experiment, ADoK-W managed to nearly rediscover the data-generating kinetic rate model; the model selected having an extra parameter in the numerator. Nevertheless, it is argued that the extra kinetic parameter is considerably smaller than the rest, which would inevitably lead to its deletion upon further investigation or model reduction.

\begin{equation}
    \hat{r}^* = \frac{9.998C_A - 4.496C_B + 0.386}{6.038C_A + 2.137C_B + 7.892}
\end{equation}

\subsection{The Decomposition of Nitrous Oxide}
Starting from the initial five experiments, ADoK-W generated and optimized a plethora of rate models, outputting the best two models (based on AIC ranking) presented below:

\begin{subequations}
\begin{gather}\label{eq:rate_decom_weak}
    \hat{r}_1(t) = \frac{(0.411C_{N_2O} - 0.062)(C_{N_2O} - 0.189) - 0.002}{C_{N_2O} - 0.189}\\
    \hat{r}_2(t) =0.411C_{N_2O} - 0.068.
\end{gather}
\end{subequations}

With these two rate models, a new experiment was suggested using MBDoE. The proposed experiment is $(C_{N_2O}(t=0), C_{N_2}(t=0), C_{O_2}(t=0)) = (0.189, 0.913, 0.926)$ M. After concatenating the measurements collected from the sixth experiment, ADoK-W was still unable to uncover the data-generating model. Instead, the methodology output the following models as the best and second-best models according to the AIC ranking:

\begin{subequations}
\begin{gather}\label{eq:rate_decom_weak_2}
    \hat{r}_1(t) = \frac{C_{N_2O}(0.927C_{N_2O} - 0.147)}{2.252C_{N_2O} + 0.046}\\
    \hat{r}_2(t) = 0.414C_{N_2O} - 0.074.
\end{gather}
\end{subequations}

With these two rate models, a new experiment was suggested using MBDoE. The proposed experiment is $(C_{N_2O}(t=0), C_{N_2}(t=0), C_{O_2}(t=0)) = (0.189, 0.913, 0.926)$ M. After concatenating the measurements collected from the seventh experiment, ADoK-S uncovers the structure and similar parameters to the data-generating rate model:

\begin{equation}
    \hat{r}^* = \frac{1.584C_{N_2O}^2}{1+3.798C_{N_2O}}.
\end{equation}

\subsection{The Hydrodealkylation of Toluene}
As presented in the main text of this work, ADoK-W rediscovered the ground truth kinetic rate model for this case study with merely the initial five experiments. The uncovered model was structurally identical to the data-generating one, whilst also having very similar kinetic values to it:

\begin{equation}
    \hat{r}^* = \frac{1.124C_TC_H}{1 + 4.932C_B + 2.928C_T}.
\end{equation}

\end{appendices}

The code used to produce all results and graphs shown in this work can be accessed at \url{https://github.com/MACServia/auto_discov_kin_rate_models}.

%%\lipsum[1-12]



\bibliographystyle{unsrtnat} 
\bibliography{references.bib}

\end{document}

%% file: abstract.tex
The industrialization of catalytic processes requires reliable kinetic models for their design, optimization and control. Mechanistic models require significant domain knowledge, while data-driven and hybrid models lack interpretability. Automated knowledge discovery methods, such as ALAMO (Automated Learning of Algebraic Models for Optimization), SINDy (Sparse Identification of Nonlinear Dynamics), and genetic programming, have gained popularity but suffer from limitations such as needing model structure assumptions, exhibiting poor scalability, and displaying sensitivity to noise. To overcome these challenges, we propose two methodological frameworks, ADoK-S and ADoK-W (Automated Discovery of Kinetic rate models using a Strong/Weak formulation of symbolic regression), for the automated generation of catalytic kinetic models using a robust criterion for model selection. We leverage genetic programming for model generation and a sequential optimization routine for model refinement. The frameworks are tested against three case studies of increasing complexity, demonstrating their ability to retrieve the underlying kinetic rate model with limited noisy data from the catalytic systems, showcasing their potential for chemical reaction engineering applications.

%% file: introduction.tex
Mathematical models are logical representations of complex phenomena, widely used in diverse fields such as physics \cite{Song_2010,Brockmann_2006}, medicine \cite{Franssen_2019,Margarit_2016}, and chemical reaction engineering \cite{Schbib_1996,Battiston_1982}. They allow researchers to distill complicated phenomena into quantitative expressions, which is essential in investigating the kinetics of a chemical system and in turn, essential in the development of industrial processes.

Models play a critical role in science and engineering, but how they are constructed remains a fundamental question. There are three classical paradigms for constructing models: mechanistic, data-driven, and hybrid modeling. Mechanistic models are derived from fundamental laws (e.g., conservation equations) but they may also include empirical expressions \cite{Baker_2018,Gernaey_2015}, and have advantages such as interpretability, extrapolatory properties, and physical meaning. However, constructing mechanistic models is time-consuming and requires domain expertise. In addition, the nonlinearity of these models can result in increased experimental effort for parameter estimation. Despite these challenges, mechanistic models are still widely established in industry and developed in research \cite{Jimenez_2011,Jedrzejewski_2014,Giessmann_2019}.

Data-driven models can be constructed quickly using only data, unlike mechanistic models that require knowledge about the system. The structure of data-driven models is flexible and can be promptly adapted to different variables or processes. They are faster to evaluate than mechanistic models \cite{Sant_Anna_2017}, making them useful in real-time simulation \cite{Zhang_2019,DelRioChanona_2018, Park_2021, Sun_2022}, optimization \cite{Petsagkourakis_2020,RioChanona_2018,Wu_2023,Natarajan_2021}, and soft sensor development \cite{Mowbray_2022,Kay_2022,Kadlec_2009}. However, since no physical knowledge is used, their extrapolatory abilities are often limited, and their performance depends on the quantity and quality of data available, which might classify their usage in certain scenarios as unsafe. %Data pre-treatment is frequently necessary before training these models.

Hybrid models aim to combine the advantages of both mechanistic and data-driven modeling. These models have a mechanistic backbone and a data-driven component that improves the fit. There are two main approaches to hybrid modeling: parallel and sequential. The parallel approach uses the data-driven block to describe the model-data mismatch, while the sequential approach uses it to describe parameters of the mechanistic backbone. With either approach, hybrid models retain the extrapolation capabilities of a mechanistic model and the flexibility and ease of construction of a data-driven model \cite{VegaRamon_2021,Mowbray_2022_1,Zhang_2020}.

Hybrid modeling offers an elegant solution to the limitations of mechanistic and data-driven modeling, albeit not the only one. Another effective approach is to use state-of-the-art statistical and machine learning methods to automatically generate and select mechanistic models using existing data. This strategy, also known as automated knowledge discovery or symbolic regression, maintains the benefits of mechanistic models while eliminating some of their drawbacks, such as the need for background knowledge and time-consuming construction \cite{Haider_2023}. The methodology presented in this work follows this paradigm.

Various methods have been proposed to solve the symbolic regression problem, including ALAMO \cite{Wilson_2017}, SINDy \cite{Brunton_2016}, \citet{Taylor_2021}, \citet{Neumann_2020}, \citet{Forster_2023}, and genetic programming \cite{Koza_1994}. However, these automated knowledge discovery frameworks face several challenges that limit their real-world applicability. Firstly, they often require structural assumptions of the underlying data-generating model, particularly non-evolutionary strategies that require a design matrix (i.e., a model library). Secondly, they may display poor scalability with respect to the number of state variables available, especially non-evolutionary strategies. Thirdly, they lack a motivated and rigorous model selection routine, and their choice of model selection routine may not be transparent or tested. Lastly, for the discovery of non-linear dynamics, they may be sensitive to noisy data when rate measurements are not directly accessible.

In this section, we introduced the importance of mathematical modeling within chemical engineering, the challenges of classical modeling paradigms, and the shortcomings of modern automated knowledge discovery methodologies. This work aims to build and benchmark two generalizable and robust methodological frameworks that integrate a rigorous model selection routine for the automated discovery of kinetic rate models. The proposed methodologies introduce two noteworthy innovations in the field of kinetic model discovery for catalytic systems. Firstly, it presents a unique approach that combines genetic programming with parameter estimation and information criteria to discover optimal state-space models for accurate rate approximation in the strong formulation. This approach contrasts with conventional methodologies where an arbitrary polynomial is typically chosen for interpolation and rate measurement estimation \cite{Forster_2023}. Additionally, the paper pioneers the application of the weak formulation of symbolic regression in genetic programming, a method that, to our knowledge, has not previously been utilized or implemented in this field. These innovations underscore the significant advancements we are contributing to the subject area.

The rest of the paper is organized as follows: in Section \ref{Methodological Frameworks} our proposed methods are motivated and described in detail; in Section \ref{Case Studies} we introduce three case studies that are used to analyze the performance of the proposed methodological frameworks; in Section \ref{Results and Discussions} the results of the study are presented and amply discussed along with the shortcomings of the proposed methodologies; and in Section \ref{Conclusions} the key findings are presented with a brief outlook on future research.

%% file: methodology.tex
We begin by briefly describing our methodologies, ADoK-S and ADoK-W (Automated Discovery of Kinetic models using a Strong/Weak formulation of symbolic regression). Both frameworks are composed of three main steps: (I) a genetic programming (GP) algorithm to facilitate candidate model generation, (II) a sequential optimization algorithm for estimating parameters of promising models, (III) and a reasoned and transparent model selection routine using the Akaike information criterion (AIC). \footnote{Selection of AIC among other criteria is explained in the `Appendix'.}

ADoK-S employs the conventional implementation of symbolic regression, or the strong formulation. This approach necessitates rate measurements for deriving rate models. However, these measurements are not experimentally accessible and need to be estimated. Following the delineated three-step procedure, ADoK-S identifies optimal concentration profiles, which describe the temporal evolution of the observed species concentrations. These profiles are then numerically differentiated to estimate the rate measurements of the reactive system. Upon rate approximation, the same three steps are carried out to discover the kinetic rate models best suited to these rates. The resultant rate model is then integrated and compared to the original concentration data.

In contrast, ADoK-W operates on the weak formulation of symbolic regression. This model proposal strategy bypasses rate estimation and constructs rate models directly from the measured concentration data. It does so by implementing the three-step process, but instead, the genetic programming algorithm contains an integration step. Consequently, the optimal rate model can be integrated and compared to the original concentration data in the same way as ADoK-S.

\begin{figure}[!htb]
    \centering
    \includegraphics[width=1\textwidth]{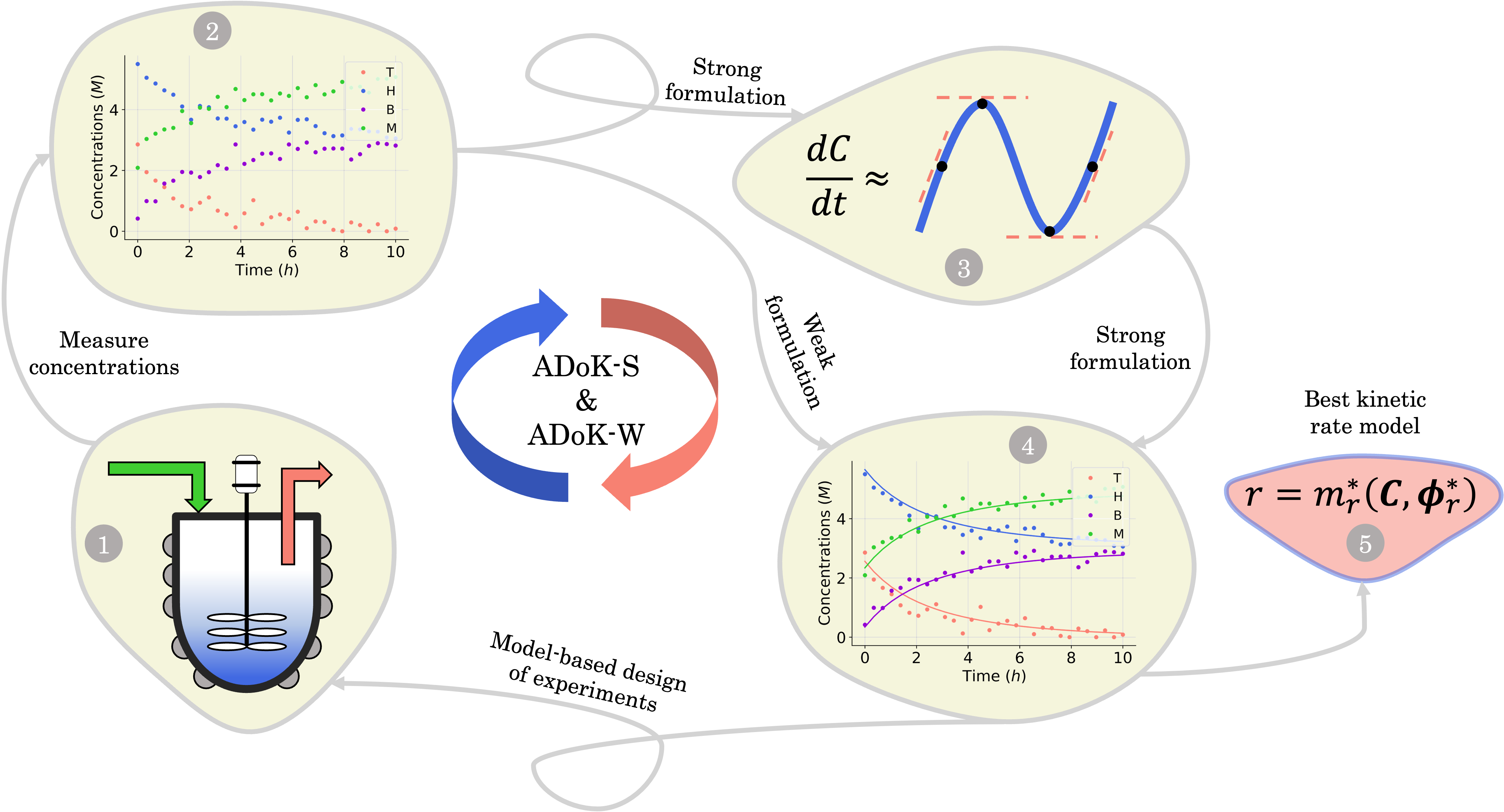}
    \caption{Flowchart representation of the ADoK-S and ADoK-W methodologies. While ADoK-S requires numerical differentiation of concentration profiles to estimate rate measurements, ADoK-W leverages an embedded integration step that enables the direct rate model extraction from concentration data. In cases of unsatisfactory model outputs, both methodologies accommodate iterative refinement using optimum experiments (determined by MBDoE) until desired accuracy or experimental budget constraints are met.}
    \label{Fig:Nice diagram}
\end{figure}

Both methodologies provide a closed-loop approach if the model output is not satisfactory, either due to violations of prior knowledge or inadequate model fitting. The modeler can choose to execute an optimum experiment tailored for the discovery task -- determined by model-based design of experiments (MBDoE) -- which can then be concatenated with the initial data set. With the new experimental data, the methodologies can be iterated and the subsequent model output examined. The number of iterations can be as many as the modeler requires or until the experimental budget is spent. Fig. \ref{Fig:Nice diagram} provides a visual representation of the workflow of both ADoK-S and ADoK-W, highlighting the most important steps and the differences between each methodology. A more detailed version is provided in Fig. \ref{fig:ADoK-S flowchart} and Fig. \ref{fig:ADoK-W flowchart}.

Here we also set the necessary mathematical notation to describe our methods precisely.
We start from the standard symbolic regression formulation \cite{Virgolin_2022}, to later introduce the weak and strong variations of our framework.

The domain set $\mathcal{Z}$ is the union of an arbitrary number of constants $\Gamma$ and a fixed number of variables $\mathcal{X}$.
The operator set $\mathcal{P}$ is the union of arithmetic operations ($\diamond: \mathbb{R}^2 \rightarrow \mathbb{R}$) and a finite set of special one-dimensional functions ($\Lambda: \mathbb{R} \rightarrow \mathbb{R}$).
The model search space $\mathcal{M}$ is the space of possible expressions to be reached by iterative function composition of the operator set $\mathcal{P}$ over the domain set $\mathcal{Z}$.

The variables can be represented as state vectors $x \in \mathbb{R}^{n_x}$.
A data point is a pair of specific states $x$ and the associated target value $y \in \mathbb{R}$ of an unknown function $f: \mathbb{R}^{n_x} \rightarrow \mathbb{R}$: $y=f(x)$.
The data set $\mathcal{D}$ consists of $n_t$ data points: $\mathcal{D} = \left\{(x^{(i)}, y^{(i)}) \mid  i = 1, \ldots, n_t \right\}$.
To quantify the discrepancy between the predictions and the target values, we can leverage any adequate positive measure function $\ell:\mathbb{R}^{n} \times \mathbb{R}^{n} \rightarrow \mathbb{R}^+$.

A symbolic model $m \in \mathcal{M}$ has a finite set of parameters $\theta_m$ whose dimension $d_m$ depends on the model.
We denote the prediction of a model under specific parameter values in functional form as $m(\cdot\mid\theta_m)$.
We use $\hat{y}_m$ to denote the prediction of a value coming from a proposed model $m$ (i.e., $\hat{y}_m = m(\cdot\mid\theta_m)$).
For our purposes, it is important to decouple the model generation step from the parameter optimization for each model.
An optimal model $m^*$ is defined as
\begin{equation} \label{eq:SR}
    m^*=\argmin_{m \in \mathcal{M}} \sum_{i=1}^{n_t}{\ell\left(\hat{y}_m^{(i)}, y^{(i)}\right)},
\end{equation}
and its optimal parameters are such that
\begin{equation} \label{eq:PE}
    \theta_{m^*}^*=\argmin_{\theta_{m^*}^*} \sum_{i=1}^{n_t}{\ell\left(\hat{y}_{m^*}^{(i)}, y^{(i)}\right)}.
\end{equation}

% Adapt the standard notation to the dynamical system formulations
In dynamical systems, the state variables are a function of time, $x(t) \in \mathbb{R}^{n_x}$, and represent the evolution of the real dynamical system within a fixed time interval $\Delta t = [t_0, t_f]$.
The dynamics are defined by the rates of change $\dot{x}(t) \in \mathbb{R}^{n_x}$ and the initial condition $x_0 = x(t=t_0)$.

% data set and sampling times
For our kinetic rate models, we assume that the $n_t$ sampling times are set within the fixed time interval, $t^{(i)} \in \Delta t$.
The concentration measurements $C$ at each time point $t^{(i)}$ are samples of the real evolution of the system $C^{(i)} \approx x(t^{(i)})$, while the rate estimates $r$ are an approximation to the rate of change $r^{(i)} \approx \dot{x}(t^{(i)})$.

Here the available data set $\mathcal{D}$ is formed by ordered pairs of time and state measurements $\mathcal{D} = \{(t^{(i)}, C^{(i)})\mid i = 1,\ldots, n_t\}$.
% predictions
As before, we use a hat to denote the prediction of either states $\hat{C}_m$ or rates $\hat{r}_m$ coming from a proposed model $m$.
The output of the models with specific parameters $\theta_m$ are denoted as $\hat{C}_m(\cdot \mid \theta_m)$ and  $\hat{r}_m(\cdot\mid\theta_m)$, respectively.
% complexity

The complexity of a model is denoted as $\mathcal{C}(m)$.\footnote{Here we use the number of nodes in an expression tree as the complexity of a symbolic expression \cite{Cranmer_2023}.}
We distinguish between families of expressions with different levels of complexity $\kappa \in \mathbb{N}$ as $\mathcal{M}^\kappa = \left\{m \in \mathcal{M} \mid \mathcal{C}(m) = \kappa\right\}$.

\subsection{Introduction to ADoK-S}\label{ADoK-S}
For ADoK-S, the objective is to find the model $m$ that best maps the states to the rates:

\begin{equation} \label{eq:strog_model}
    \hat{r}_m(t \mid \theta_m) = m(x(t) \mid \theta_m).
\end{equation}

For this to be done directly, an estimation of the rates of change $r^{(i)}$ must be derived from the available concentration measurements $C^{(i)}$. To solve this, our approach forms an intermediate symbolic model $\eta$ such that $\eta(t^{(i)}) \approx C^{(i)}$ following the standard symbolic regression procedure, described in \eqref{eq:SR} and \eqref{eq:PE}, with our model selection process described in Section \ref{Model Selection}.

Since this model is differentiable, its derivatives provide an approximation to the desired rates: $\dot{\eta}\left(t^{(i)}\right) \approx r^{(i)}$. With these estimated values available, the optimization problem can be written as follows. The outer level optimizes over model proposals for a fixed level of complexity $\kappa$,

\begin{align}\label{Eq:1.1}
    m^\star = \argmin_{m \in \mathcal{M^\kappa}} \sum_{i=1}^{n_t} \ell\left(\hat r_m(t^{(i)}\mid\theta_m), r^{(i)} \right),
\end{align}

while the inner level optimizes over the best model's parameters,

\begin{align}\label{eq:PE Strong}
    \theta_{m^\star}^\star = \argmin_{\theta_{m^\star}} \sum_{i=1}^{n_t} \ell\left(\hat r_{m^\star}(t^{(i)}\mid\theta_{m^\star}), r^{(i)} \right).
\end{align}

\begin{figure}[!htb]
    \centering
    \includegraphics[width=0.7\textwidth]{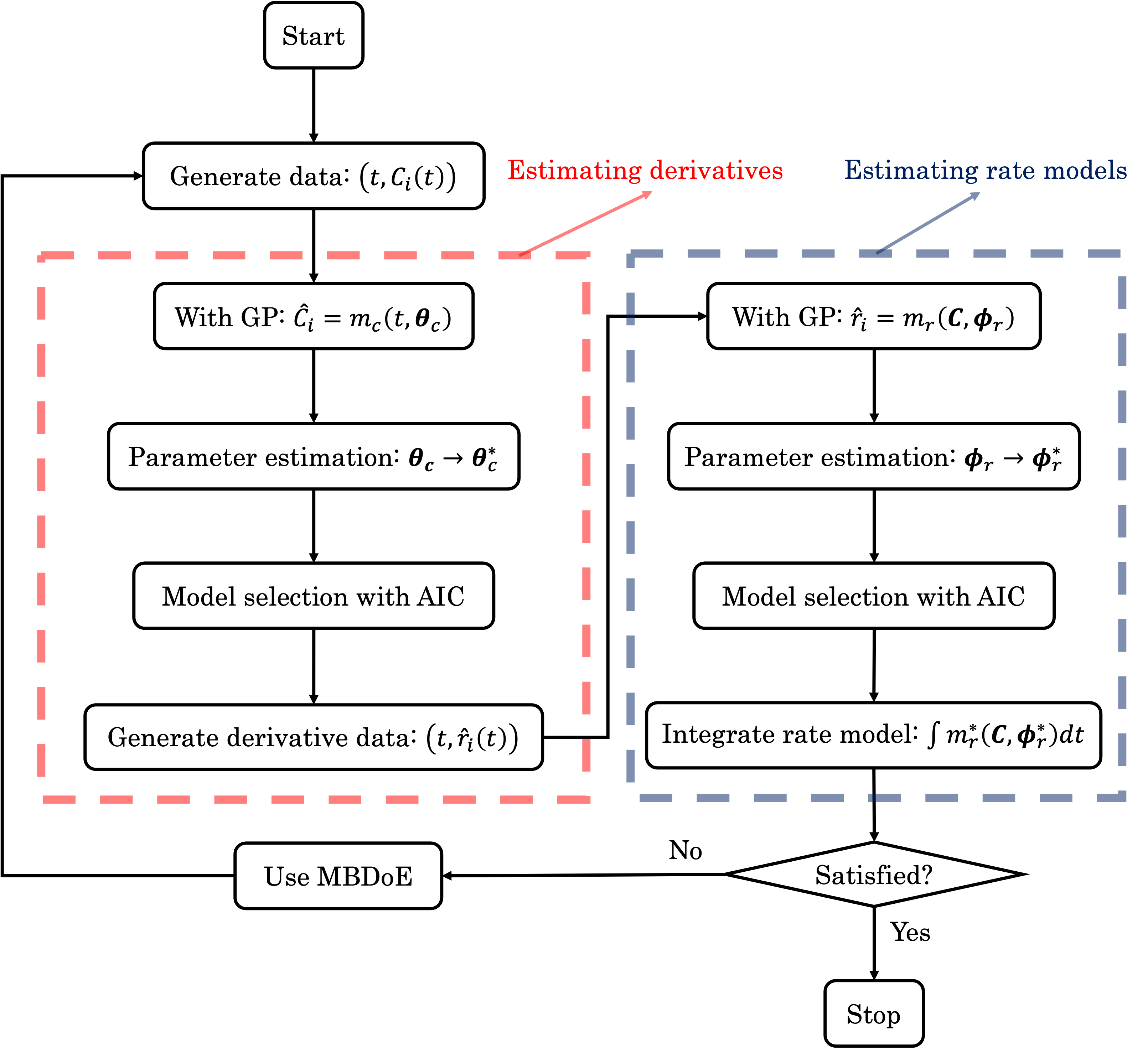}
    \caption{The flowchart of ADoK-S (\textbf{A}utomated \textbf{D}iscovery \textbf{o}f \textbf{K}inetics using a \textbf{S}trong formulation of symbolic regression); the \textcolor{red}{red} and \textcolor{blue}{blue} dashed boxes represent the steps where rate measurements and rate models are estimated, respectively.}
    \label{fig:ADoK-S flowchart}
\end{figure}

In Eq. \eqref{Eq:1.1} and \eqref{eq:PE Strong}, $\ell$ represents the residual sum of squares (RSS). The whole process of this approach is showcased in Fig. \ref{fig:ADoK-S flowchart}.

\subsection{Introduction to ADoK-W}
For ADoK-W, we aim to find the model $m$ that best maps state variables to the differential equation system that define the state dynamics to then predict the concentration evolution:

\begin{subequations}
\begin{equation} \label{eq:weak_model}
    \dot x_m(t \mid \theta_m) = m(x(t)\mid \theta_m),
\end{equation}

\begin{equation}
    \hat C_m(t\mid \theta_m) = C_0 + \int_{t_0}^{t} \dot x_m(\tau \mid \theta_m) d\tau,
\end{equation}
\end{subequations}

where the initial condition $C_0$ is the first concentration measurement. For this formulation, the outer level optimizes over model proposals for a specific complexity level $\kappa$ as well,

\begin{align}\label{Eq:1.2}
    m^\star = \argmin_{m \in \mathcal{M^\kappa}} \sum_{i=1}^{n_t} \ell\left(\hat C_m(t^{(i)}\mid\theta_m), C^{(i)} \right),
\end{align}

while the inner level optimizes over the parameters of the best model,

\begin{align}\label{Eq:1.3}
    \theta_{m^\star}^\star = \argmin_{\theta_{m^\star}} \sum_{i=1}^{n_t} \ell\left(\hat C_{m^\star}(t^{(i)}\mid\theta_{m^\star}), C^{(i)} \right).
\end{align}

\begin{figure}[!htb]
    \centering
    \includegraphics[width=0.6\textwidth]{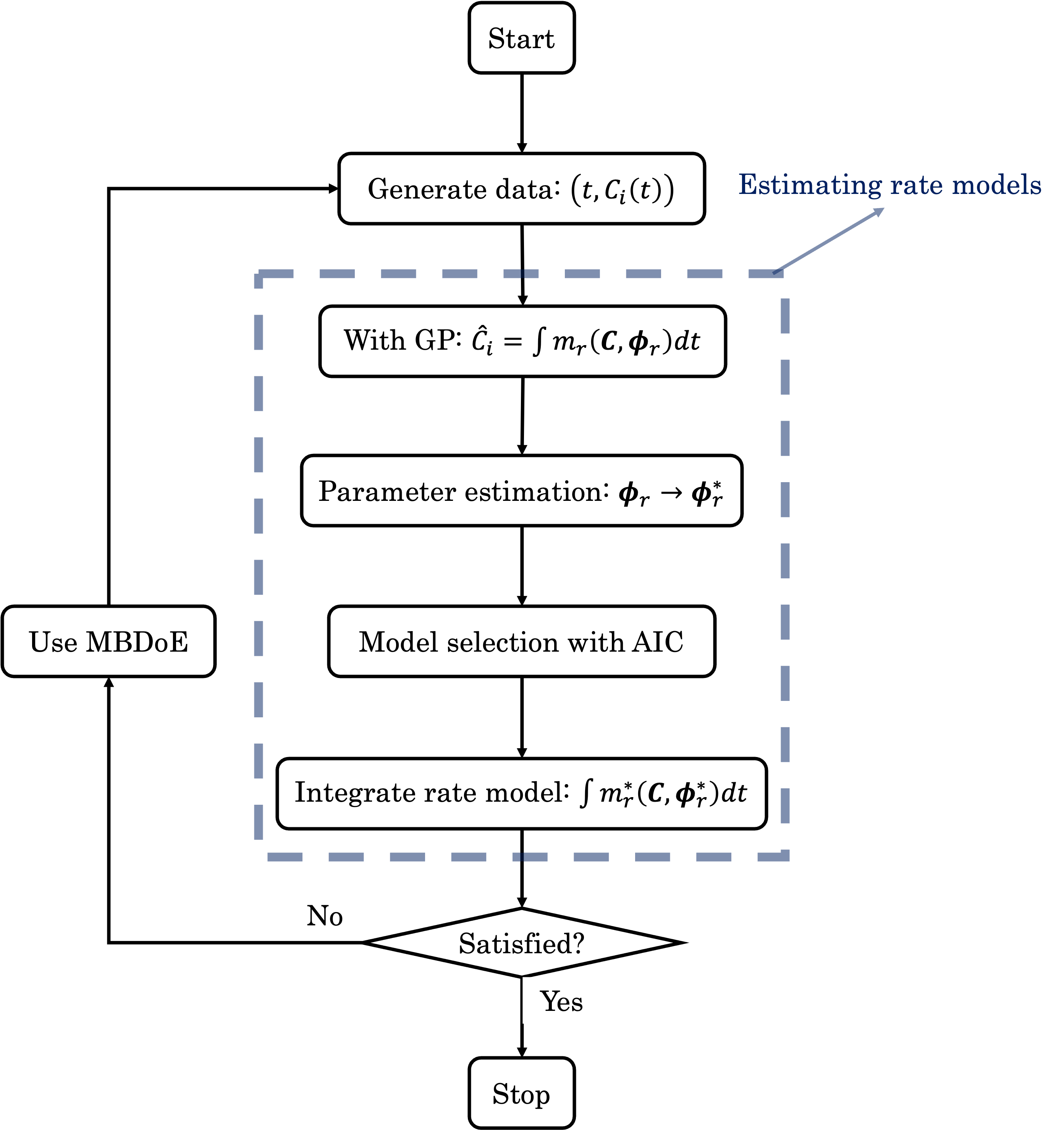}
    \caption{The flowchart of ADoK-W (\textbf{A}utomated \textbf{D}iscovery \textbf{o}f \textbf{K}inetics using a \textbf{W}eak formulation of symbolic regression); the \textcolor{blue}{blue} dashed box represent the steps where rate models are estimated.}
    \label{fig:ADoK-W flowchart}
\end{figure}

In Eq. \eqref{Eq:1.2} and \eqref{Eq:1.3}, $\ell$ represents the RSS. The whole process of this variation is depicted in Fig. \ref{fig:ADoK-W flowchart}.

\subsection{Model Selection}\label{Model Selection}

Given a model $m$ with parameters $\theta_m$ of dimension $d_m$, the Akaike information criterion is defined as

\begin{equation} \label{eq:AIC}
    \text{AIC}_m = 2\mathcal{L}(\mathbf{\theta}_m\mid\mathcal{D}) + 2d_m, 
\end{equation}

where $\mathcal{L}$ represents specifically the negative log-likelihood (NLL). Given two competing models, $m_1$ and $m_2$, the preferred model would be the one with the lowest AIC value calculated by Eq. \eqref{eq:AIC}. The choice of AIC for model selection within the ADoK-S and ADoK-W framework is motivated in detail in the `Appendix'.

\subsection{Model-Based Design of Experiments}
It is possible that the data set used for the regression is not enough to provide an adequate model proposal. For this scenario, and under the assumption that the experimental budget is not fully spent, it is possible to leverage the implicit insights in the optimized models to extract an informative proposal for a new experiment. For this purpose, we may search for an initial condition which maximizes the discrepancy between state predictions $\hat x(t)$ of the best two proposed models, $\eta$ and $\mu$, using the available data set. This MBDoE approach was developed in \cite{Hunter_1965}:

\begin{equation}\label{Eq: MBDoE}
     x_0^{(new)} = \argmax_{x_0} \int_{t_0}^{t_f} \ell\left(\hat x_\eta \left(\tau\mid\theta_\eta^\star \right), \hat x_\mu \left(\tau\mid\theta_\mu^\star \right) \right)\, d\tau.
\end{equation}

In Eq. \eqref{Eq: MBDoE}, $\ell$ represents the RSS. Starting from the proposed initial condition, an experiment can be carried out to obtain a new batch of data points to be added to the original data set. Finally, the whole process of model proposal and selection can be redone with the enhanced data set, closing the loop between informative experiments and optimal models.

%% file: casestudies.tex
To assess the efficacy of our methodologies, we undertook an analysis of three case studies of catalytic reactions drawn from the literature: a theoretical isomerization reaction \cite{Marin_2019}, the decomposition of nitrous oxide \cite{Levenspiel_1998}, and the hydrodealkylation of toluene \cite{Fogler_2016}. For conciseness, our discussion focuses primarily on the most complex example -- the hydrodealkylation of toluene. A comprehensive discussion on this case, along with the others, is available in the `Appendix'.

\subsection{The Hydrodealkylation of Toluene}\label{Hydrodealkylation}
The hydrodealkylation of toluene reaction can be represented by Eq. \eqref{Eq:2.1} while Eq. \eqref{Eq:2.2} provides a description of the reaction rate \cite{Fogler_2016}, where the kinetic parameters were defined as: $K_A=2$ M$^{-1}$ h$^{-1}$, $K_B=9$ M$^{-1}$ and $K_C=5$ M$^{-1}$. 

\begin{gather}\label{Eq:2.1}
    C_6H_5CH_3 + H_2 \rightleftharpoons C_6H_6 + CH_4
\end{gather}

\begin{gather}\label{Eq:2.2}
    r=-\frac{dC_{T}}{dt}=-\frac{dC_{H}}{dt}=\frac{dC_{B}}{dt}=\frac{dC_{M}}{dt} =\frac{K_AC_TC_H}{1+K_BC_B+K_CC_T}
\end{gather}

In Eq. \eqref{Eq:2.2}, \textit{T}, \textit{H}, \textit{B}, and \textit{M} correspond to toluene, hydrogen gas, benzene and methane, respectively. Starting from Eq. \ref{Eq:2.2}, an in-silico data set is established wherein $\Delta t = [0, 10]$ h and $n_t = 30$. This data set is the composed of five different experiments, each ran at different initial conditions; these experiments were randomly picked from a $2^k$ factorial design \cite{Mee_2009}. 

For all experiments, the system is assumed to be both isochoric and isothermal, and Gaussian noise is added to replicate experimental data. The generated data set for the second and fourth experiments are presented in Fig. \ref{Fig:Results ADoK-S} a) and e). The data set, providing 150 datapoints, has a realistic size for kinetic studies \cite{Schrecker_2023, Waldron_2020, Taylor_2021}, especially considering recent advancements in high-throughput setups. 

%% file: results_n_discussion.tex
\subsection{ADoK-S Performance}\label{ADoK-S Performance}
As outlined in Fig. \ref{fig:ADoK-S flowchart}, the first stage in deriving kinetic models from dynamic concentration trajectories in ADoK-S is proposing concentration profile models. This necessitates the application of a GP algorithm (in ADoK-S we use the implementation from \citet{Cranmer_2023}), featuring the following expression construction rules: $\mathcal{P} = \{+,-,\div,\times,\exp\}$ and $\mathcal{X} = \{t\}$, where \textit{t} denotes the time variable. This selection is considered reasonable based on our physical understanding of kinetic modeling -- a clear route of injecting expert knowledge into the symbolic search. 

It is important to note that at times, the solution to the fundamental ordinary differential equation (ODE) system, delineating the kinetics of the reactive system, may not exist as a closed-form expression. In these cases, any proposed concentration model given any construction rules will be flawed. Nonetheless, in most tested cases (see `Appendix'), the chosen construction rules have evidenced their capability to successfully approximate the behavior of the concentration trajectories, and more importantly, the rate measurements, regardless of the existence of a closed-form expression. 

\begin{figure}[htb!]
    \centering
    \includegraphics[width=0.9\textwidth]{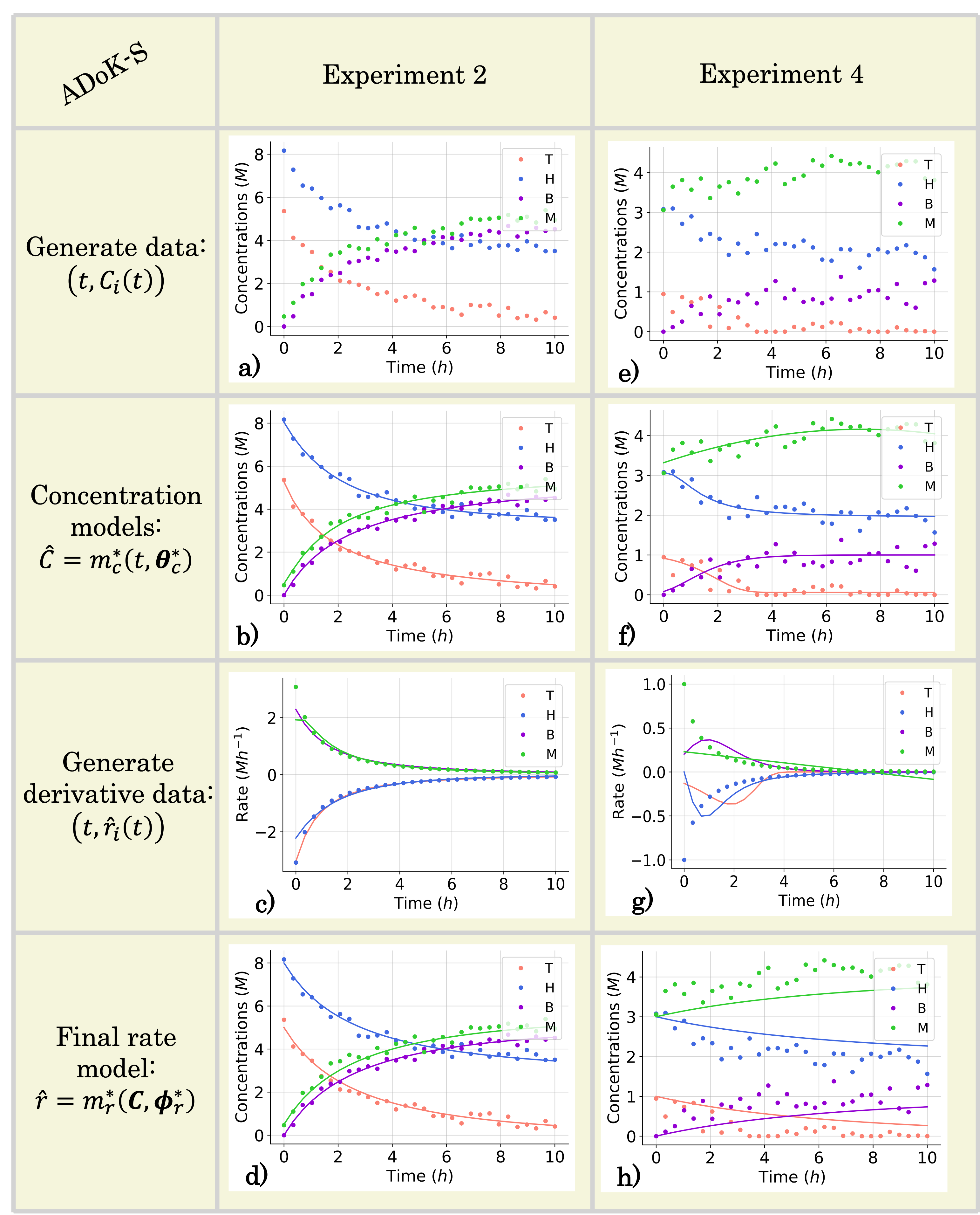}
    \caption{The conditions for the second and fourth computational experiment are $(C_{T,0}, C_{H,0}, C_{B,0}, C_{M,0}) \in {(5, 8, 0, 0.5), (1, 3, 0, 3)}$ M, respectively, where \textit{T}, \textit{H}, \textit{B}, \textit{M} denote toluene, hydrogen, benzene and methane, respectively. \textbf{a)} and \textbf{e)}: the measured concentration data for the second and fourth experiments which are used in the execution of ADoK-S for the hydrodealkylation of toluene. \textbf{b)} and \textbf{f)}: the concentration profiles selected by AIC that model the dynamic trajectories of the observable species' concentrations in the second and fourth experiments as a function of time. These models are used to approximate the rate measurements. \textbf{c)} and \textbf{g)}: numerical derivatives of the selected concentration profiles and the true rate measurements (which realistically are inaccessible). \textbf{d)} and \textbf{h)}: response of the selected rate model after the first iteration of the ADoK-S with the initial set of experiments for the second and fourth experiments.}
    \label{Fig:Results ADoK-S}
\end{figure}

Fig. \ref{Fig:Results ADoK-S} b) and c) demonstrate ADoK-S' ability to approximate the concentration profile as well as the rate measurements of a reactive system. However, Fig. \ref{Fig:Results ADoK-S} f) and g) shows the opposite, where the ADoK-S clearly struggles to capture the appropriate behavior of the dynamic evolution of the concentrations, which is subsequently translated to poor rate approximations. These results further motivate the development of ADoK-W and incentivize its use in complicated case studies despite its longer computational time. 

In this particular case study, we construct four concentration models for each experiment -- specifically, ($\hat{C}_{T,i}$, $\hat{C}_{H,i}$, $\hat{C}_{B,i}$, $\hat{C}_{M,i}$) for $i \in [1, 2, ..., 5]$, where $i$ denotes the experiment number. It is important to underscore that the development of each of these models is an autonomous process. Some might contend that this methodology could result in models that violate essential physical principles such as the conservation of mass. However, it is argued that the primary objective at this initial phase is to approximate the system's rate measurements accurately, thereby facilitating the creation of precise kinetic models. Therefore, in this context, a certain level of physical inconsistency might be tolerable.

This section focuses on the results from the second experiment, as the same methodology has been employed across all other experiments. The GP algorithm proposes model structures for the concentrations of \textit{T}, \textit{H}, \textit{B} and \textit{M} for each complexity level, which is capped by the user. We present below the proposed concentration profiles for \textit{T} in the second experiment. Here $p_i$ represents the $i^{\textrm{th}}$ parameter that can be estimated from the time-dependent concentration data set for a specific model. Further, $\hat{C}_{i}$ denotes the $i^{\textrm{th}}$ proposed concentration model of species \textit{T} in the second experiment by ADoK-S.

\begin{subequations}
\begin{gather}\label{Eq:2.3}
    \hat{C}_1(t) = p_1 \\
    \hat{C}_2(t) = \frac{p_1}{\exp{(t)}} \\
    \hat{C}_3(t) = \frac{p_1}{p_2 + t} \\
    \hat{C}_4(t) = \frac{p_1}{\exp{(p_2)} + t} \\
    \hat{C}_5(t) = \frac{p_1}{t + p_2} - p_3 \\
    \hat{C}_6(t) = \frac{p_1}{\exp{(p_2t)} + t} \\
    \hat{C}_7(t) = \frac{p_1}{p_2 + \frac{t}{p_3}} - p_4
\end{gather}
\end{subequations}

After the generation of the concentration model structures, the next step involves parameter estimation. This is aimed at finding the optimal values that minimize the error between the response of the concentration models and the measured concentrations. 

With parameter values determined, both the negative log-likelihood (NLL) function and the Akaike information criterion (AIC) (Eq. \eqref{eq:AIC}) are computed for each model to enable the model selection process. From the proposed models, $m_7$ is the chosen model to approximate the rate of consumption measurements for the second experiment for species \textit{T}.

Fig. \ref{Fig:Results ADoK-S} b) and f) presents the concentration models developed, optimized, and selected through ADoK-S for all species in the second and fourth experiment. Following the selection of concentration models, the generation rates (for products) and consumption rates (for reactants) are estimated via numerical differentiation of these models. Fig. \ref{Fig:Results ADoK-S} c) and g) showcases these estimated rates over time, in comparison with the rate measurements from the real system $\dot x(t)$, generally inaccessible in practice. As mentioned, the methodology excels in the rate estimation for experiment two, but struggles in doing so for experiment four. The early equilibrium ($\sim$ 2 hours), combined with high additive noise, renders experiment four less kinetically informative, making it challenging for frameworks like ADoK-S to extract meaningful insights and approximate rate measurements.

In alignment with the flowchart presented in Fig. \ref{fig:ADoK-S flowchart}, the next stage of ADoK-S employs the same GP algorithm (used to derive concentration profiles) to propose rate models. This procedure unfolds iteratively, refining populations of rate models with the aim to satisfy Eq. \eqref{Eq:1.1}. The rules for constructing expressions are $\mathcal{P} = \{+,-,\div,\times\}$ and $\mathcal{X} = \{C_T,C_H,C_B,C_M\}$, a selection based on our prior understanding of kinetic models -- yet another route to inject expert knowledge into the methodology. Based on these construction rules, the GP algorithm suggests 13 rate model structures; for the sake of brevity, we present a select few:

\begin{subequations}
\begin{gather}\label{eq:rates}
    \hat{r}_1 = K_1 \\
    \hat{r}_2 = K_1C_T \\
    \hat{r}_3 = K_1C_TC_H \\
    \hat{r}_4 = K_1C_H - K_2C_T \\
    \hat{r}_5 = K_1C_TC_H - K_2C_T \\
    \hat{r}_6 = \frac{K_1C_T((C_H - K_2)(K_3 - C_B) + K_4)}{C_B - K_5}.
\end{gather}
\end{subequations}
 
The parameters $K_i$ for $i \in [1,2,...,5]$ of the dynamical models are estimated from the concentration data, a process known as dynamic parameter estimation. The parameter estimation problem is solved by satisfying Eq. \eqref{eq:PE Strong}, utilizing the ABC and LBFGS optimization algorithms to identify the optimal solution. Upon computing the NLL and AIC values for all proposed models, the selected model is $\hat{r}_3$, and its response is presented in Fig. \ref{Fig:Results ADoK-S} d) and h). 

None of the equations shown in Eq. \eqref{eq:rates}, including $\hat{r}_3$, match the data-generating rate model shown in Eq. \eqref{Eq:2.2}. Additionally, as displayed in \ref{Fig:Results ADoK-S} h), the model's response is in poor agreement with the concentration data from the fourth experiment. Therefore, ADoK-S must undergo another iteration using MBDoE. For the MBDoE, the top two models yielded by ADoK-S, namely $\hat{r}_3$ and $\hat{r}_5$, are used to propose a discriminatory experiment by solving Eq. \eqref{Eq: MBDoE}.

The MBDoE procedure suggests running a sixth experiment with initial conditions $(C_{T,0}, C_{H,0}, C_{B,0}, C_{M,0}) = (1.948, 7.503, 1.232, 2.504)$ M. The new experiment undergoes the same sequence of operations as the initial five: generate, optimize, and select the best concentration models to approximate the rates. Once the rates of the new experiment are computed, they are concatenated with the prior approximations, and new rate models are accordingly generated, optimized, and selected. For the sake of brevity, the proposed concentration and rate models are not presented here, but the best ($\hat{r}_1$) and second-best ($\hat{r}_2$) kinetic models chosen by ADoK-S following the addition of the extra experiment are presented below:

\begin{subequations}
\begin{gather}\label{Eq:2.5}
    \hat{r}_1 = \frac{C_TC_H}{C_T + K_1} \\
    \hat{r}_2 = K_1C_T(-K_2C_B^2 + K_3C_B + K_4C_H - C_T + K_5) + K_6.
\end{gather}
\end{subequations}

Although the fitness of the new model improved compared to the initially selected rate model, the fitness is still not satisfactory. As such, ADoK-S undergoes one more iteration where the MBDoE procedure suggests running a seventh experiment with initial conditions $(C_{T,0}, C_{H,0}, C_{B,0}, C_{M,0}) = (2.560, 5.654, 0.341, 2.337)$ M. The kinetic model selected by ADoK-S after the seventh experiment, denoted as $r^*$, is presented below:

\begin{gather}\label{Eq:2.6}
    r^* = \frac{K_1C_TC_H}{K_2 + K_3C_B + K_4C_T}. 
\end{gather}

As demonstrated, after two iterations of ADoK-S, the methodology is able to uncover a structurally identical kinetic model (Eq. \eqref{Eq:2.6}) to the data-generating one (Eq. \eqref{Eq:2.2}).

\clearpage

\subsection{ADoK-W Performance}
As exposed in Section \ref{ADoK-S Performance}, ADoK-S demonstrates limitation in approximating rate measurements for complex systems under conditions fraught with noise, as anticipated by \cite{Bertsimas_2023}. This motivates the desirability of circumventing rate estimations for knowledge discovery where possible.

The theoretical shortcomings of ADoK-S, when combined with its suboptimal performance in discerning the ground-truth model underpinning hydrodealkylation of toluene, sparked the creation of ADoK-W. ADoK-W, exploiting the weak formulation of symbolic regression, mitigates the need for rate approximations in proposing rate models. This novel design allows ADoK-W to suggest rate models directly from concentration data, as opposed to limiting model proposals to direct input-output mappings. This innovation lies in incorporating an integration step within the genetic programming algorithm tasked with model proposal.

Nonetheless, beyond this variation, ADoK-W operates in identical fashion to ADoK-S. Initially, models are formulated using genetic programming. The most optimal models in each complexity category are optimized by solving a parameter estimation problem. From the refined model set, the one with the lowest AIC value is selected. Should the modeler find the algorithm output unsatisfactory (and the experimental budget allows), additional experiments guided by MBDoE can be conducted, the measurements concatenated to the previous data set, and another iteration of the ADoK-W algorithm may be executed.

The rules established for rate model construction for ADoK-S remain consistent in the execution of ADoK-W. By solving Eq. \eqref{Eq:1.2} at different complexity levels, the genetic programming algorithm formulated six rate models, presented below:

\begin{subequations}
\begin{gather}\label{Eq:2.7}
    \hat{r}_1 = K_1 \\
    \hat{r}_2 = K_1C_T \\
    \hat{r}_3 = K_1C_TC_H \\
    \hat{r}_4 = \frac{K_1C_TC_H}{C_B + K_2} \\
    \hat{r}_5 = \frac{K_1C_TC_H}{C_T + K_2C_B} \\
    \hat{r}_6 = \frac{K_1C_TC_H}{K_2 + K_3C_T + K_4C_B}.
\end{gather}
\end{subequations}

\begin{figure}[htb!]
    \centering
    \includegraphics[width=0.9\textwidth]{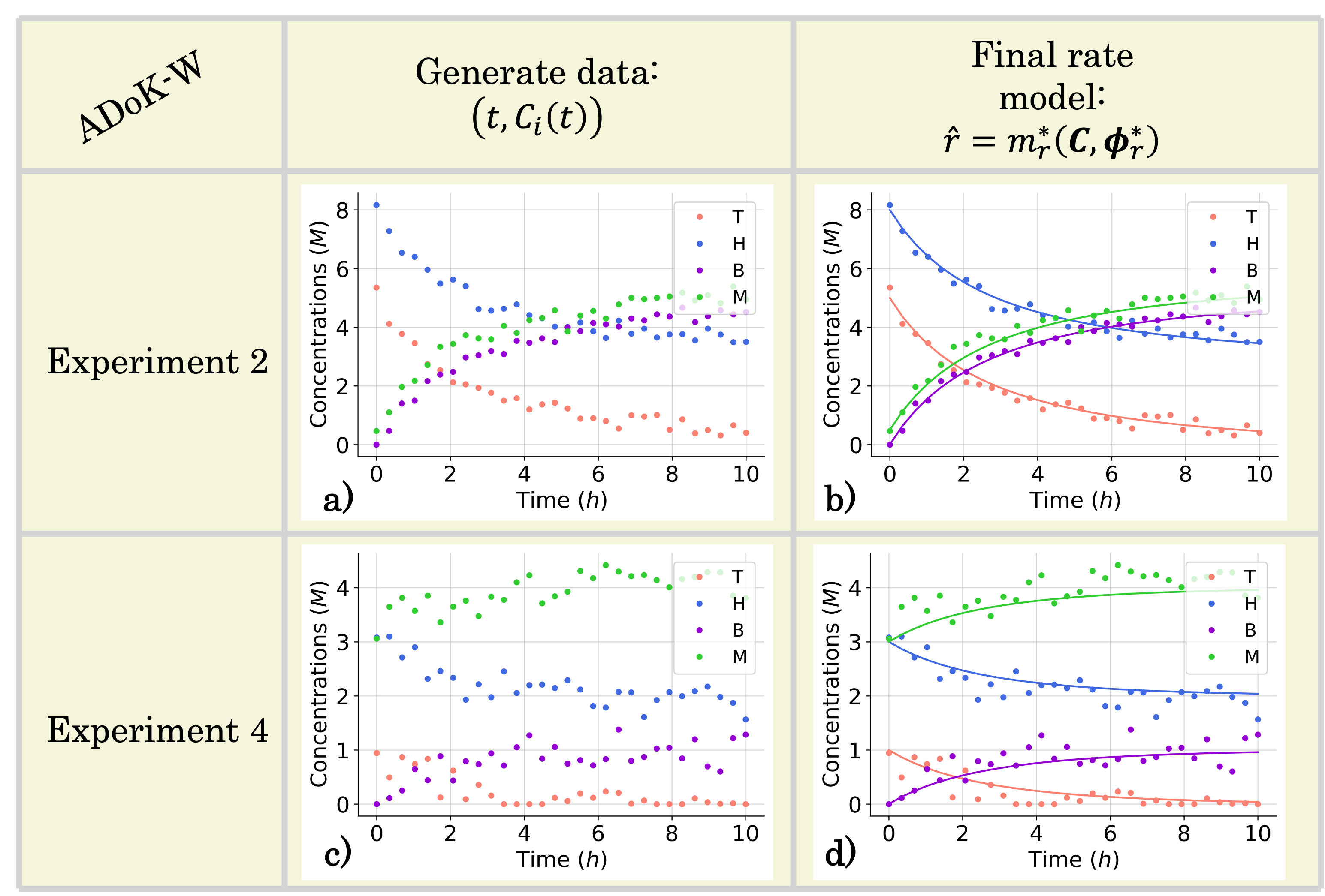}
    \caption{The conditions for the second and fourth computational experiment are  $(C_{T,0}, C_{H,0}, C_{B,0}, C_{M,0}) \in {(5, 8, 0, 0.5), (1, 3, 0, 3)}$ M, respectively, where \textit{T}, \textit{H}, \textit{B}, \textit{M} denote toluene, hydrogen, benzene and methane, respectively. \textbf{a)} and \textbf{c)}: the measured concentration data for the second and fourth experiments which are used in the execution of ADoK-W for the hydrodealkylation of toluene. \textbf{b)} and \textbf{d)}: response of the selected rate model after the first and only iteration of the ADoK-W for the second and fourth experiments.}
    \label{Fig:Results ADoK-W}
\end{figure}

Following the framework delineated in Fig. \ref{fig:ADoK-W flowchart}, upon the proposition of the best rate models, these models are refined by identifying the parameters that satisfy Eq. \eqref{Eq:1.3}. Following optimization, AIC values are computed, and the model with the lowest value is chosen. For this case study, the selected model, $r_6$, coincides exactly with the ground-truth model in Eq. \eqref{Eq:2.2} without any MBDoE iterations of the methodology. Fig. \ref{Fig:Results ADoK-W} shows the measured concentration data and the response of the selected kinetic model by ADoK-W for the second and fourth experiments. 

%% file: conclusions.tex
In this work, we introduce two data-driven frameworks, ADoK-S and ADoK-W, which tackle the symbolic regression problem to discover kinetic rate models from noisy concentration measurements. Using a genetic programming algorithm coupled with parameter estimation and an information criterion, these methods generate, refine, and select models without undue restrictions. Unlike black-box and hybrid models that may obscure interpretability, or traditional mechanistic models that could be time and resource intensive to construct, our methods offer a transparent and efficient process for interpretable model development.

While ADoK-S necessitates rate measurements to propose rate models, in line with a strong formulation, ADoK-W bypasses this need, directly creating rate models from concentration data, a characteristic of a weak formulation. In the case study of hydrodealkylation of toluene, both methods successfully identified the underlying rate model of the reaction.

However, due to errors in rate approximations and system complexity, ADoK-S required two extra iterations to discover the ground-truth kinetic model. On the other hand, ADoK-W found the data-generating model using solely the initial five experiments, showing better performance in complex spaces, albeit without ‘free lunch’. Where ADoK-S can propose rate models within minutes, ADoK-W requires hours to do the same. 

The results from this case study -- along with the ones presented in the `Appendix', which demonstrate similar success in uncovering kinetic rate models from data -- highlight the potential of using automated knowledge discovery methods in kinetic model development in reaction engineering and catalysis. The summarized results are presented in Table \ref{Table: Summarized Results}. While we demonstrated this potential with minimal prior knowledge, the long computation times hint at the need for integrating physical constraints, like the law of conservation of mass and equilibrium behavior, to reduce the search space and improve computational efficiency.

\begin{table}[htb!]
\caption{The summarized results of the performance of ADoK-S and ADoK-S against all three case studies explored.}
\begin{tabular}{p{0.22\textwidth}p{0.22\textwidth}p{0.22\textwidth}p{0.22\textwidth}}
\hline
    & Hypothetical isomerization reaction &  Decomposition of nitrous oxide & Hydrodealkylation of toluene \\
\hline
    Number of iterations -- ADoK-S & 2 & 2 & 3\\

    Number of iterations -- ADoK-W & 2 & 3 & 1\\

    Data-generating kinetic model & $\frac{7C_A-3C_B}{4C_A+2C_B+6}$ & $\frac{2C_{N_2O}^2}{1+5C_{N_2O}}$ & $\frac{2C_TC_H}{1+9C_B+5C_T}$ \\

    Rate model uncovered -- ADoK-S & $\frac{8.666C_A-3.642C_B}{4.976C_A+2.525C_B+7.003}$ & $\frac{1.937C_{N_2O}^2}{1+4.803C_{N_2O}}$ & $\frac{1.669C_TC_H}{1+7.347C_B+4.439C_T}$ \\

    Rate model uncovered -- ADoK-W & $\frac{9.998C_A - 4.496C_B + 0.386}{6.038C_A + 2.137C_B + 7.892}$ & $\frac{1.584C_{N_2O}^2}{1+3.798C_{N_2O}}$ & $\frac{1.124C_TC_H}{1 + 4.932C_B + 2.928C_T}$\\
% \hline
%     Mean squared error -- ADoK-S & 0.338 & 0.104 & 0.160 \\
% \hline
%     Mean squared error -- ADoK-W & 0.444 & 0.110 & 0.142 \\
\hline
\end{tabular}
\label{Table: Summarized Results}
    \centering
\end{table}

It is important to mention that the success of any data-driven approach, including the ones presented here, depends heavily on the data used. The data assumptions made in this case study may not always hold true. For fast reactions or reactions with different species, the sampling rate and assuming that all species can be measured might be unrealistic. The assumption of perfect device calibration and no systematic errors, although optimistic, may not always be an accurate representation of real experimental setups.

Lastly, while ADoK-S and ADoK-W have been designed for discovering kinetic models of catalytic systems, they can be extended without major modifications to explore the dynamics of non-reactive systems. This broadens their potential applications to other fields and disciplines.